\documentclass[journal]{IEEEtran}

\IEEEoverridecommandlockouts

\usepackage{cite}
\usepackage{amsmath,amssymb,amsfonts}
\usepackage{algorithmic}
\usepackage{graphicx}
\usepackage{textcomp}
\def\BibTeX{{\rm B\kern-.05em{\sc i\kern-.025em b}\kern-.08em
    T\kern-.1667em\lower.7ex\hbox{E}\kern-.125emX}}

\usepackage{amsmath}
\usepackage{amsfonts}
\usepackage{color}
\usepackage{soul}
\usepackage{cite}
\usepackage{graphicx}
\usepackage{algorithm}
\usepackage{mathtools}
\usepackage{bm}
\usepackage{booktabs}
\usepackage{textcomp}

\usepackage{pgf}
\usepackage{printlen}

\usepackage{hyperref}

\newtheorem{theorem}{Theorem}
\newtheorem{corollary}{Corollary}
\newtheorem{assumption}{Assumption}

\newtheorem{proposition}{Proposition}

\newenvironment{proof}{\paragraph*{Proof}}{\hfill$\square$}

\DeclareMathSymbol{\shortminus}{\mathbin}{AMSa}{"39}

\DeclareMathOperator*{\argmin}{\mathrm{argmin}}
\DeclareMathOperator*{\rank}{\mathrm{rank}}
\DeclareMathOperator*{\nnz}{\mathrm{nnz}}

\newcommand{\real}{\mathbb{R}}

\newif\ifreviewermode
\reviewermodefalse 

\usepackage[dvipsnames]{xcolor}
\usepackage{marginnote}
\usepackage{xstring}
\newcommand{\revres}[2]{%
    \ifreviewermode
        \IfEqCase{#1}{%
            {1}{\textcolor{brown}}%
            {2}{\textcolor{magenta}}%
            {3}{\textcolor{olive}}%
            {4}{\textcolor{teal}}%
            {5}{\textcolor{LimeGreen}}%
            {6}{\textcolor{violet}}%
        }[\PackageError{revres}{Undefined revres option: #1}{}]{#2}%
    \else
        {#2}%
    \fi
}%
\newcommand{\revmargin}[3][false]{%
    \ifreviewermode
        \def\placemarginnote{%
            \IfEqCase{#2}{%
                {1}{\textcolor{brown}}%
                {2}{\textcolor{magenta}}%
                {3}{\textcolor{olive}}%
                {4}{\textcolor{teal}}%
                {5}{\textcolor{LimeGreen}}%
                {6}{\textcolor{violet}}%
            }[\PackageError{revmargin}{Undefined revmargin option: #2}{}]%
            {\marginnote{#3}}%
        }%
        \IfStrEq{#1}{true}{%
            {
                \reversemarginpar
                \placemarginnote
                \normalmarginpar 
            }%
        }{%
            \placemarginnote
        }%
    \else
    \fi
}

\usepackage[normalem]{ulem}
\newcommand{\remove}[1]{%
  \ifreviewermode
    \textcolor{red}{\sout{#1}}
  \else
  \fi
}%

\begin{document}

\title{Sparsity-Promoting Reachability Analysis and \\Optimization of Constrained Zonotopes}

\author{Joshua A. Robbins, Jacob A. Siefert, and Herschel C. Pangborn
\thanks{Joshua A. Robbins and Herschel C. Pangborn are with the Department of Mechanical Engineering, The Pennsylvania State University, University Park, PA 16802 USA (e-mail: {\tt\small jrobbins@psu.edu, hcpangborn@psu.edu}).}
\thanks{Jacob A. Siefert is with the Penn State Applied Research Laboratory, University Park, PA 16802 USA (e-mail: {\tt\small jas7031@psu.edu}).}
}

\maketitle

\begin{abstract}
The constrained zonotope is a polytopic set representation widely used for set-based analysis and control of dynamic systems. This paper develops methods to formulate and solve optimization problems for dynamic systems in real time using reachability analysis with constrained zonotopes.
An alternating direction method of multipliers (ADMM) algorithm is presented that makes efficient use of the constrained zonotope structure. To increase the efficiency of the ADMM iterations, reachability calculations are presented that increase the sparsity of the matrices used to define a constrained zonotope when compared to typical methods.
The developed methods are used to formulate and solve predictive control, state estimation, and safety verification problems. Numerical results show that optimization times using the proposed approach are competitive with state-of-the-art QP solvers and conventional problem formulations.
A combined set-valued state estimation and moving horizon estimation algorithm is presented and experimentally demonstrated in the context of robot localization.
\end{abstract}

\section{Introduction}
Novel set-theoretic methods continue to be developed as important tools for analyzing and controlling dynamic systems. For example, sets can be used to rigorously bound the reachable states of a system or to define constraints to which a controller must adhere. 
\revmargin{6}{6.1}\revres{6}{Traditional set representations such as halfspace representation (H-rep) polytopes, vertex representation (V-rep) polytopes, ellipsoids, and intervals remain widely used. More modern offerings include include AH-polytopes~\cite{sadraddini2019linear}, ellipsotopes~\cite{kousik2022ellipsotopes}, constrained convex generators~\cite{silvestre2021constrained},} \revmargin{5}{5.2}\revres{5}{and zonotopic sets (i.e., zonotopes and generalizations of zonotopes).
Zonotopic sets} have become particularly popular\remove{ set representations}, as evidenced by their appearance in several open-source toolboxes \cite{koeln2023zonolab, althoff2016cora, vinod2024pycvxset,  bogomolov2019juliareach, michaux2023rdf, hadjiloizou2022Formal, schupp2022recent, rego2025zeta}. Key to the appeal of these set representations is that they have closed-form expressions for many important set operations and\remove{, in the context of reachability analysis,}\revmargin{5}{5.2} have favorable complexity growth when compared to traditional set representations 
\remove{such as halfspace representation (H-rep) polytopes, vertex representation (V-rep) polytopes, and ellipsoids}
\cite{althoff2021set}. 

These attributes make \revmargin{5}{5.2}\revres{5}{zonotopic sets} attractive candidates for use in real-time reachability analysis. In~\cite{althoff2014online}, for example, zonotope-based reachability analysis was used to verify the safety of system trajectories in autonomous driving experiments. In addition to safety verification, reachability analysis can be used to formulate optimal control and planning problems~\cite{seo2022real, liu2024refine, bravo2006robust, schurmann2018reachset}.

This paper focuses on constrained zonotopes, which can represent any convex polytope and--critically--are closed under generalized intersections \cite{scott2016constrained}. 
Constrained zonotopes have seen extensive application to problems in set-valued state estimation (SVSE) \cite{scott2016constrained,  raimondo2016closed, rego2020guaranteed}, robust controllable set \cite{vinod2025projection} and robust positively invariant set~\cite{raghuraman2022set} computation, hierarchical control \cite{koeln2020vertical}, and tube-based model predictive control (MPC)~\cite{andrade2024tube}, to name a few. 

While it is typically very efficient to perform operations on constrained zonotopes, analyzing these sets often requires the use of numerical optimization algorithms. For example, assessing whether a constrained zonotope contains a point requires solving a linear program (LP) feasibility problem \cite{scott2016constrained}. Similarly, assessing whether a constrained zonotope is empty also requires solving a feasibility problem. By contrast, point containment for H-rep polytopes only requires matrix-vector multiplication and inequality evaluation. 
Despite this fundamental reliance on numerical optimization, the problem of constructing these sets to be efficiently analyzed by specific optimization algorithms has received little attention.

\begin{figure}[t]
    \centering
    \input{figs/zono_opt_conceptual.pgf}
    \caption{Online reachability analysis with constrained zonotopes is used to construct the feasible space of an optimal control or estimation problem. ADMM is then used to find a trajectory within the reachable sets that minimizes a specified quadratic objective function.}
    \label{fig:conceptual-figure}
\end{figure}

\subsection{Gaps in the Literature}
Despite the favorable complexity growth of constrained zonotopes when compared to more traditional set representations, sets with high complexity, e.g., those defined by hundreds or thousands of constraints and generator variables, are often produced in practice~\cite{vinod2025projection, zhang2022safety, rego2020guaranteed}. High complexity sets may have large memory requirements and generally take longer to analyze using numerical optimization algorithms than lower complexity sets. This challenge has motivated work on complexity reduction techniques~\cite{raghuraman2022set, scott2016constrained, rego2025novel}. Complexity is reduced by finding sets that are equivalent to or approximate the original set, but for which the matrices and vectors used to define the set have lower dimension. Sparse matrices can also be used to reduce the memory demands of a constrained zonotope representation, as noted in~\cite{raghuraman2022set}. The sparsity patterns of these matrices have not been explicitly considered, however.

To complement complexity reduction techniques, optimization algorithms that scale well with large numbers of states and constraints and exploit sparsity in the problem data can be used to reduce the computational burden. 
First-order optimization methods are known to scale better to large problem sizes than interior point or active set methods \cite{stellato2020osqp}. One such method is the alternating direction method of multipliers (ADMM). A key advantage of ADMM is that it can be formulated to exploit particular problem structure \cite{boyd2011distributed}. For example, distributed ADMM formulations have been widely applied in machine learning contexts \cite{boyd2011distributed, huang2019dp, elgabli2020q}. Specializations of ADMM for control include a parallelized MPC optimization routine~\cite{rey2020admm} and a general method for formulating two degree-of-freedom controllers~\cite{srikanthan2023augmented}.

In the specific contexts of motion planning and moving horizon estimation, the particular properties and structure of hybrid zonotopes\revres{1}{---an extension of constrained zonotopes that includes integer variables---}\revmargin{1}{1.5}are exploited within a mixed-integer optimization routine in \cite{robbins2024efficient, robbins2024mixed, thompson2025mixed}. While some of the optimization techniques presented in these papers apply to constrained zonotopes, the methods are restricted to multi-stage problem structures and problems with diagonal cost matrices. 
\revmargin{5}{5.2}\remove{Efficient solution of more general optimization problems formulated using constrained zonotopes has not been considered.}

\subsection{Contributions}
This paper considers the real-time formulation and solution of constrained optimization problems for dynamic systems. Our approach is illustrated in Fig.~\ref{fig:conceptual-figure}. The feasible space of a constrained optimal control or estimation problem for a discrete-time dynamic system is constructed using reachability analysis with constrained zonotopes, and then ADMM is used to find the optimal trajectory within the reachable sets.
Expressing the feasible space of an optimization problem through set-theoretic operations provides a unique framework from which to consider its structure and computational complexity. This leads to new insights in constructing a problem representation that is amenable to efficient optimization, and in selecting an optimization algorithm to most efficiently exploit the problem representation.
We pursue computational efficiencies in both the reachability and optimization steps to facilitate their real-time execution.

The key contributions of this paper are as follows: \revmargin[true]{5}{5.2} \revres{5}{\begin{enumerate}
    \item Novel reachability calculations for affine dynamics using constrained zonotopes decrease the number of non-zero elements in the matrices that define the set when compared to existing methods. 
    \item An ADMM formulation makes efficient use of the structure of constraint sets represented as constrained zonotopes when solving convex quadratic programs (QPs).
    \item A method is proposed for efficiently verifying that a constrained zonotope is empty using the ADMM iterations.
    \item An algorithm based on reachability analysis of constrained zonotopes combines moving horizon estimation (MHE) and SVSE. The effectiveness of this algorithm is demonstrated experimentally.
\end{enumerate}}

The new methods proposed in this paper are available in an open-source toolbox\revres{4}{\footnote{\url{https://github.com/psu-PAC-Lab/ZonoOpt}}}\revmargin{4}{4.14}, implemented in C++ with Python bindings, that facilitates reachability calculations using constrained zonotopes and sparse linear algebra with tailored optimization routines.

\subsection{Outline}
The remainder of this paper is organized as follows. Preliminary information is given in Sec.~\ref{sec:preliminaries}. In Sec.~\ref{sec:sparsity}, reachability calculations for linear systems using constrained zonotopes are presented that result in matrices with high sparsity. Sec.~\ref{sec:ADMM} proposes an ADMM algorithm that is tailored to constrained zonotopes. In Sec.~\ref{sec:applications}, the sparse reachability calculations and ADMM algorithm are applied to MPC, MHE, and safety verification problems. Numerical results for these problems are given in Sec.~\ref{sec:numerical-examples}. 
Using the proposed approach based on reachability analysis with sparse constrained zonotopes and ADMM, the MPC problems are solved faster than when using state-of-the-art QP solvers and a commonly used sparse MPC problem formulation, especially for large problem sizes. The SVSE/MHE example illustrates an equivalence between reachability analysis and MHE, as a set-valued state estimate is extracted directly from the MHE problem formulation. The infeasibility certificate is used for efficient safety verification of system trajectories in a third numerical example. Sec.~\ref{sec:experiment} experimentally demonstrates the real-time capabilities of the proposed methods, with the combined SVSE/MHE algorithm applied for online robot localization.
Sec.~\ref{sec:conclusion} concludes the paper.
\section{Preliminaries}  \label{sec:preliminaries}

\subsection{Notation} Unless otherwise stated, scalars are denoted by lowercase letters, vectors by boldface lowercase letters, matrices by uppercase letters, and sets by calligraphic letters. 
$\mathbf{0} = \begin{bmatrix} 0 & \cdots & 0 \end{bmatrix}^T$ and $\mathbf{1} = \begin{bmatrix} 1 & \cdots & 1 \end{bmatrix}^T$ denote vectors consisting entirely of zeroes and ones, respectively, of appropriate dimensions. $I$ denotes the identity matrix. The operator $\nnz{(\cdot)}$ gives the number of non-zero elements of a sparse matrix. 
For a set $\mathcal{C}$ and a function $f(\mathbf{x})$, $\min_{\mathbf{x} \in \mathcal{C}} f(\mathbf{x})$ and $\max_{\mathbf{x} \in \mathcal{C}} f(\mathbf{x})$ denote the minimum and maximum values of $f(\mathbf{x})$ such that $\mathbf{x} \in \mathcal{C}$. The indicator function for a set $\mathcal{C}$ is defined as $$I_{\mathcal{C}}(\mathbf{x}) = \begin{cases} {0, \; \text{ if } \mathbf{x} \in \mathcal{C}} \;, \\ {\infty, \; \text{otherwise}} \;. \end{cases}$$

A \emph{closed interval} $[x]$ is defined by its lower bound  $\underline{x}$ and upper bound $\overline{x}$, as $[x] = [\underline{x}, \overline{x}]$. A \emph{box} $[\mathbf{x}]$ is a vector of closed intervals such that $[\mathbf{x}] = \begin{bmatrix} [x_1] & [x_2] & \cdots\end{bmatrix}^T$.

\subsection{Zonotopes and Constrained Zonotopes}

A set $\mathcal{Z}$ is a zonotope if $\exists G \in \real^{n \times n_G}$ and $\mathbf{c} \in \real^n$ such that~\cite{ziegler2012lectures}
\begin{equation}
    \mathcal{Z} = \left\{ G \bm{\xi} + \mathbf{c} \middle| \bm{\xi} \in [-1,1]^{n_G} \right\} \;.
\end{equation}
Zonotopes are centrally symmetric convex polytopes.

Constrained zonotopes extend the definition of zonotopes to include equality constraints. A set $\mathcal{Z}$ is a constrained zonotope if there exists a generator matrix $G \in \real^{n \times n_G}$, center $\mathbf{c} \in \real^n$, constraint matrix $A \in \real^{n_C \times n_G}$, and constraint vector $\mathbf{b} \in \real^{n_C}$ such that
\begin{equation} \label{eq:conzono-definition}
    \mathcal{Z} = \left\{ G \bm{\xi} + \mathbf{c} \middle| A \bm{\xi} = \mathbf{b},\; \bm{\xi} \in [-1,1]^{n_G} \right\} \;.
\end{equation}
Here, $n$ is the dimension of the constrained zonotope, $n_G$ is the number of generators, and $n_C$ is the number of equality constraints. The shorthand notation $\mathcal{Z} = \left\langle G, \mathbf{c}, A, \mathbf{b} \right\rangle$ is often used in place of \eqref{eq:conzono-definition}. Constrained zonotopes can represent any convex polytope \cite{scott2016constrained}.

One major advantage of constrained zonotopes when compared to other polytopic set representations is that they have closed form expressions for key set operations,
\begin{subequations} \label{eq:set-ops}
\begin{align}
&R \mathcal{Z} + \mathbf{s} = \left\langle RG, R\mathbf{c} + \mathbf{s}, A, \mathbf{b} \right\rangle \;, \label{eq:set-ops-lin-map} \\
&\mathcal{Z}_1 \times \mathcal{Z}_2 = \left\langle \begin{bmatrix}
    G_1 & 0 \\
    0 & G_2
\end{bmatrix}, \begin{bmatrix}
    \mathbf{c}_1 \\ \mathbf{c}_2
\end{bmatrix}, \begin{bmatrix}
    A_1 & 0 \\
    0 & A_2
\end{bmatrix}, \begin{bmatrix}
    \mathbf{b}_1 \\ \mathbf{b}_2
\end{bmatrix} \right\rangle \;, \label{eq:set-ops-cart-prod} \\
&\mathcal{Z}_1 \oplus \mathcal{Z}_2 = \left\langle \begin{bmatrix}
    G_1 & G_2
\end{bmatrix}, \mathbf{c}_1 + \mathbf{c}_2, \begin{bmatrix}
    A_1 & 0 \\
    0 & A_2
\end{bmatrix}, \begin{bmatrix}
    \mathbf{b}_1 \\ \mathbf{b}_2
\end{bmatrix} \right\rangle \;, \label{eq:set-ops-mink-sum} \\
&\mathcal{Z}_1 \cap_R \mathcal{Z}_2 = \left\langle \begin{bmatrix}
    G_1 & 0
\end{bmatrix}, \mathbf{c}_1, \right. \nonumber \\
&\qquad \qquad \qquad \qquad \left. \begin{bmatrix}
    A_1 & 0 \\
    0 & A_2 \\
    R G_1 & -G_2
\end{bmatrix}, \begin{bmatrix}
    \mathbf{b}_1 \\ \mathbf{b}_2 \\ \mathbf{c}_2 - R \mathbf{c}_1
\end{bmatrix} \right\rangle \;, \label{eq:set-ops-intersection}
\end{align}
\end{subequations}
where \eqref{eq:set-ops-lin-map} is the affine map \revres{6}{$\{R\mathbf{z} + \mathbf{s} | \mathbf{z} \in \mathcal{Z}\}$}, \eqref{eq:set-ops-cart-prod} is the Cartesian product \revres{6}{$\{ [\mathbf{z}_1^T~\mathbf{z}_2^T]^T | \mathbf{z}_1 \in \mathcal{Z}_1,\; \mathbf{z}_2 \in \mathcal{Z}_2 \}$}, \eqref{eq:set-ops-mink-sum} is the Minkowski sum \revres{6}{$\{\mathbf{z}_1 + \mathbf{z}_2 | \mathbf{z}_1 \in \mathcal{Z}_1,\; \mathbf{z}_2 \in \mathcal{Z}_2 \} $}, and \eqref{eq:set-ops-intersection} is the generalized intersection, \revmargin{6}{6.2}\revres{6}{$\{\mathbf{z} \in \mathcal{Z}_1 | R \mathbf{z} \in \mathcal{Z}_2\}$.}

\subsection{Alternating Direction Method of Multipliers}
ADMM solves optimization problems of the form \cite{boyd2011distributed}
\begin{subequations} \label{eq:admm-general}
\begin{align}
    &\min_{\mathbf{x}, \mathbf{z}} f(\mathbf{x}) + g(\mathbf{z})\;, \\
    &\text{s.t.}\; A\mathbf{x} + B \mathbf{z} = \mathbf{c} \;, \label{eq:admm-general-eq-cons}
\end{align}
\end{subequations}
\revres{6}{where $\mathbf{x} \in \real^n$, $\mathbf{z} \in \real^m$, and $\mathbf{c} \in \real^p$, and $f : \real^n \rightarrow \real \cup \{ \infty \}$ and $g : \real^m \rightarrow \real \cup \{ \infty \}$}\revmargin{6}{6.2}

The algorithm is defined by the iterations
\begin{subequations} \label{eq:admm-iterations-general}
\begin{align}
    &\mathbf{x}_{k+1} = \argmin_{\mathbf{x}}(f(\mathbf{x}) + \frac{\rho}{2} ||A \mathbf{x} + B \mathbf{z}_k - \mathbf{c} + \mathbf{u}_k||_2^2) \;, \label{eq:admm-gen-a} \\
    &\mathbf{z}_{k+1} = \argmin_{\mathbf{z}} (g(\mathbf{z}) + \frac{\rho}{2} ||A \mathbf{x}_{k+1} + B \mathbf{z} - \mathbf{c} + \mathbf{u}_k||_2^2) \;, \label{eq:admm-gen-b} \\
    &\mathbf{u}_{k+1} = \mathbf{u}_k + A \mathbf{x}_{k+1} + B \mathbf{z}_{k+1} - \mathbf{c} \;. \label{eq:admm-gen-c}
\end{align}
\end{subequations}
where $\rho>0$ is the ADMM penalty parameter which weighs satisfaction of the constraint \eqref{eq:admm-general-eq-cons}.

\revres{6}{The convergence of~\eqref{eq:admm-iterations-general} to feasible and optimal $\mathbf{x}^*$, $\mathbf{z}^*$, and $\mathbf{u}^*$ is guaranteed when 1) $f$ and $g$ are closed, proper, and convex, and 2) the augmented Lagrangian 
\begin{equation}
    \mathcal{L} = f(\mathbf{x}) + g(\mathbf{z}) + \frac{\rho}{2} ||A \mathbf{x} + B \mathbf{z} - \mathbf{c} + \mathbf{u}||_2^2 \;,
\end{equation}
has a saddle point~\cite{boyd2011distributed}.}\revmargin{6}{6.2}

Typical stopping criteria are to monitor the primal and dual residuals $\mathbf{r}_p \in \real^p$ and $\mathbf{r}_d \in \real^n$ for convergence, where
\begin{equation}
    \mathbf{r}_p = A \mathbf{x}_k + B \mathbf{z}_k - \mathbf{c} \;, \quad \mathbf{r}_d = \rho A^T B (\mathbf{z}_{k+1} - \mathbf{z}_k) \;.
\end{equation}
Common absolute convergence criteria are $||\mathbf{r}_p||_2 < \sqrt{p} \epsilon_p$ and $||\mathbf{r}_d||_2 < \sqrt{n} \epsilon_d$ with $\epsilon_p>0$ and $\epsilon_d>0$ the primal and dual convergence tolerances, respectively \cite{boyd2011distributed}.

\subsection{Interval Arithmetic}
The following interval arithmetic operations are used in this paper \cite{jaulin2001interval}:
\begin{subequations}
\begin{align}
    &[x] + [y] = [\underline{x} + \underline{y}, \overline{x} + \overline{y}] \;, \\
    &[x] - [y] = [\underline{x} - \overline{y}, \overline{x} - \underline{y}] \;, \\
    &\alpha [x] = \begin{cases}
        [\alpha \underline{x}, \alpha \overline{x}], \; \text{if } \alpha \geq 0 \;, \\
        [\alpha \overline{x}, \alpha \underline{x}], \; \text{if } \alpha < 0 \;,
    \end{cases} \\
    &[x][y] = [\min{(\underline{x}\underline{y}, \underline{x}\overline{y}, \overline{x}\underline{y}, \overline{x}\overline{y})}, \max{(\underline{x}\underline{y}, \underline{x}\overline{y}, \overline{x}\underline{y}, \overline{x}\overline{y})}] \;.
\end{align}
\end{subequations}

The product $\mathbf{v}^T [\mathbf{x}]$ is derived from the above arithmetic equations as 
\begin{equation}
    \mathbf{v}^T [\mathbf{x}] = \sum_{i} v_i [x_i] \;.
\end{equation}

A scalar $x \in [x]$ if and only if $\underline{x} \leq x \leq \overline{x}$. Similarly, a vector $\mathbf{x} \in [\mathbf{x}]$ if and only if $x_i \in [x_i] \; \forall i$. 
\section{Reachability with Constrained Zonotopes via Sparsity-Promoting Operations} \label{sec:sparsity}
For optimization algorithms that rely on sparse linear algebra, the computational complexity of optimizing over or assessing the feasibility of a constrained zonotope will depend on the number of non-zero elements of the matrices involved \cite{demmel1997applied}. However, previous assessments of the computational complexity of set operations on constrained zonotopes have assumed dense linear algebra \cite{althoff2021set}. The following presents closed-form set operations on constrained zonotopes \remove{to compute reachable sets}\revres{3}{for estimation and control applications}\revmargin{3}{3.1} that result in matrices with significantly fewer non-zero elements than existing methods.

Consider the linear system 
\begin{subequations} \label{eq:linsys}
\begin{align}
    &\mathbf{x}_{k+1} = A \mathbf{x}_k + B \mathbf{u}_k \;, \\
    &\text{s.t.} \; \forall k: \mathbf{x}_k, \mathbf{x}_{k+1} \in \mathcal{S} \;,\; \mathbf{u}_k \in \mathcal{U} \;,
\end{align}
\end{subequations}
where $\mathcal{S}$ and $\mathcal{U}$ are the state and input \emph{domain sets} of the system dynamics, respectively~\cite{siefert2025reachability}. 
The state domain set $\mathcal{S}$ captures limits on the system states. This is particularly useful \revres{3}{in SVSE and}\revmargin{3}{3.1}
when using reachability analysis to build MPC and MHE problems.\remove{, as demonstrated in Secs.~\ref{sec:applications-mpc} and \ref{sec:app-mhe}.}\revmargin{5}{5.3}
\revres{3}{In applications where there is no \textit{a priori} bound $\mathcal{S}$ on the system states, intersections are not necessary and reachability analysis with zonotopes can be used.}\revmargin{3}{3.1}

\subsection{Methods for Computing the N-Step Reachable Set}
We next present three different methods for computing the N-step reachable set of the system in~\eqref{eq:linsys}, each of which achieves a different sparsity in the matrices defining the reachable set.

\subsubsection{Standard Method}
Given a set of initial states $\mathcal{X}_0$, the $N$-step reachable set $\mathcal{X}_N$ can be computed by the iterations
\begin{equation} \label{eq:reachability-trad}
    \mathcal{X}_{k+1} = \left(A \mathcal{X}_k \oplus B \mathcal{U} \right) \cap \revres{1}{\mathcal{S}}\revmargin{1}{1.6} \;, \forall k\in \{0, ..., N-1\} \;.
\end{equation}
Eq.~\ref{eq:reachability-trad} is analogous to the SVSE equations provided in~\cite{scott2016constrained}.

\remove{Constrained zonotopes are a popular choice for computing $\mathcal{X}_N$ because all of the operations in \eqref{eq:reachability-trad} have closed-form expressions as given in \eqref{eq:set-ops}. Reachability analysis using constrained zonotopes is known to have favorable complexity growth when compared to methods based on other polytopic set representations
. However, these analyses do not consider the sparsity structure of the resulting sets.}
\revmargin[true]{5}{5.3}\remove{Indeed the sparsity of the $G$ and $A$ matrices in the $N$-step reachable set $\mathcal{X}_N$ can be improved significantly, as will be demonstrated in Secs.~\ref{sec:sparse-graph} and \ref{sec:sparse-ours}.}

Consider $\mathcal{X}_k = \left\langle G_{x,k}, \cdot, A_{x,k}, \cdot \right\rangle$, $\mathcal{U} = \left\langle G_u, \cdot, A_u, \cdot \right\rangle$, and $\mathcal{S} = \left\langle G_s, \cdot, A_s, \cdot \right\rangle$. Vector terms are assumed dense and have been omitted for conciseness.
The generator and constraint matrices for the iterates of \eqref{eq:reachability-trad} are explicitly given as 
\begin{subequations} \label{eq:reachability-trad-conzono}
\begin{align}
    &G_{x,k+1} = \begin{bmatrix}
        A G_{x,k} & B G_u & 0
    \end{bmatrix} \;, \label{eq:reachability-trad-conzono-G} \\
    &A_{x,k+1} = \begin{bmatrix}
        A_{x,k} & 0 & 0 \\
        0 & A_u & 0 \\
        0 & 0 & A_s \\
        A G_{x,k} & B G_u & -G_s
    \end{bmatrix} \;. \label{eq:reachability-trad-conzono-A}
\end{align}
\end{subequations}

\remove{From inspection of \eqref{eq:reachability-trad-conzono}, the sparsity of matrices $G_N$ and $A_N$ in $\mathcal{X}_N = \left\langle G_N, \cdot, A_N, \cdot \right\rangle$ scales as $\nnz{(G_N)} \propto N$ and $\nnz{(A_N)} \propto N^2$.} 

\subsubsection{Graph of Function Method} \label{sec:sparse-graph}
In~\cite{siefert2025reachability}, a reachability analysis technique using graphs of functions was proposed. The graph of the function $\psi(\cdot)$ is defined as 
\revmargin{1}{1.7}
\begin{equation}
    \Psi = \left\{ \begin{bmatrix}
        \mathbf{p} & \mathbf{q}
    \end{bmatrix}^T \middle| \mathbf{p} \in \mathcal{D}, \mathbf{q} \revres{6}{=} \psi(\revres{1}{\mathbf{p}}), \revres{6}{\mathbf{q}} \subseteq \mathcal{Q} \right\} \;, 
\end{equation}
\revmargin{6}{6.3}where $\mathcal{D}$ is the domain set of the inputs $\mathbf{p}$ and $\mathcal{Q}$ is the domain set of the outputs $\mathbf{q}$. While that work targeted nonlinear and hybrid systems, graphs of functions can be used to perform reachability calculations for affine systems as well. For the system in \eqref{eq:linsys}, the graph of the function $\psi(\mathbf{x}, \mathbf{u}) = A \mathbf{x} + B \mathbf{u}$ with $\mathcal{D} = \mathcal{S} \times \mathcal{U}$ and $\mathcal{Q} = \mathcal{S}$ is given as
\begin{equation}
    \Psi = \left( \begin{bmatrix}
        I & 0 \\
        0 & I \\
        A & B
    \end{bmatrix} (\mathcal{S} \times \mathcal{U}) \right) \cap_{\begin{bmatrix}
        0 & 0 & I
    \end{bmatrix}} \mathcal{S} \;. 
\end{equation}
Using $\Psi$, the $N$-step reachable set can be computed via the iterations 
\begin{equation} \label{eq:reachability-state-update}
    \mathcal{X}_{k+1} = \begin{bmatrix} 0 & 0 & I \end{bmatrix} \left( \Psi \cap_{\begin{bmatrix} I & 0 & 0 \\ 0 & I & 0 \end{bmatrix}} (\mathcal{X}_k \times \mathcal{U}) \right) \;.
\end{equation}

The constrained zonotope $\Psi$ is given explicitly as
\begin{equation} \label{eq:Psi-explicit}
    \Psi = \left\langle 
    \begin{bmatrix}
        G_s & 0 & 0 \\
        0 & G_u & 0 \\
        A G_s & B G_u & 0 
    \end{bmatrix}, \cdot,    
    \begin{bmatrix}
        A_s & 0 & 0 \\
        0 & A_u & 0 \\
        0 & 0 & A_s \\
        A G_s & B G_u & -G_s
    \end{bmatrix}, \cdot
    \right\rangle \;.
\end{equation}
Defining the rows of the generator matrix in \eqref{eq:Psi-explicit} to be \revres{6}{$G_{\psi, 1}$, $G_{\psi, 2}$, and $G_{\psi, 3}$} and the columns of the constraint matrix to be \revres{6}{$A_{\psi,1}$, $A_{\psi,2}$, and $A_{\psi,3}$}\revmargin{6}{6.3}, the generator and constraint matrices for the iterates \eqref{eq:reachability-state-update} are given explicitly as 
\begin{subequations} \label{eq:reachability-state-update-conzono}
\begin{align}
    &G_{x,k+1} = \begin{bmatrix}
        \revres{6}{G_{\psi,3}} & 0 & 0
    \end{bmatrix} \;, \\
    &A_{x,k+1} = \begin{bmatrix}
        \revres{6}{A_{\psi,1}} & \revres{6}{A_{\psi,2}} & \revres{6}{A_{\psi,3}} & 0 & 0 \\
        0 & 0 & 0 & A_{x,k} & 0 \\
        0 & 0 & 0 & 0 & A_u \\
        \revres{6}{G_{\psi,1}} & 0 & 0 & -G_{x,k} & 0 \\
        0 & \revres{6}{G_{\psi,2}} & 0 & 0 & -G_{u}
    \end{bmatrix} \;.
\end{align}
\end{subequations}

\remove{In contrast with \eqref{eq:reachability-trad-conzono}, the $N$-step reachable set $\mathcal{X}_N$ computed using \eqref{eq:reachability-state-update-conzono} has constant $\nnz{(G_N)}$ with respect to $N$ and $\nnz{(A_N)} \propto N$.}

\subsubsection{Sparsity-Promoting Method} \label{sec:sparse-ours}
Next we present a method for computing $\mathcal{X}_N$ that produces a constrained zonotope with $G$ and $A$ matrices containing fewer non-zeros than either \eqref{eq:reachability-trad} or \eqref{eq:reachability-state-update}. Further, the number of generators and constraints used to represent $\mathcal{X}_N$ is identical to \eqref{eq:reachability-trad}.

Consider the iterations
\begin{equation} \label{eq:reachability-alt-int}
  \mathcal{X}_{k+1} = \begin{bmatrix} 0 & 0 & I \end{bmatrix} \left( (\mathcal{X}_k \times \mathcal{U} \times S) \cap_{\begin{bmatrix} A & B & -I \end{bmatrix}} \{\mathbf{0}\} \right) \;.
\end{equation}

\begin{proposition} \label{prop:intersection-reachability}
Eq.~\eqref{eq:reachability-alt-int} gives the one-step reachable set of system~\eqref{eq:linsys}.
\begin{proof}
    Define the right-hand side of \eqref{eq:reachability-alt-int} to be $\mathcal{Z}$. By applying set operation definitions, $\mathcal{Z}$ is given as $\mathcal{Z} = \left\{ \mathbf{c} \middle| \begin{bmatrix} \mathbf{a}^T & \mathbf{b}^T & \mathbf{c}^T \end{bmatrix}^T \in (\mathcal{X}_k \times \mathcal{U} \times{\mathcal{S}}), A \mathbf{a} + B \mathbf{b} = \mathbf{c} \right\}$, and by definition $\mathcal{Z} = \mathcal{X}_{k+1}$.
\end{proof}
\end{proposition}

The generator and constraint matrices for the iterates of \eqref{eq:reachability-alt-int} are given explicitly as
\begin{subequations} \label{eq:reachability-alt-int-conzono}
\begin{align}
    &G_{x,k+1} = \begin{bmatrix}
        0 & 0 & G_s
    \end{bmatrix} \;, \\ 
    &A_{x,k+1} = \begin{bmatrix}
        A_{x,k} & 0 & 0 \\
        0 & A_u & 0 \\
        0 & 0 & A_s \\
        AG_{x,k} & BG_u & -G_s
    \end{bmatrix} \;.
\end{align}
\end{subequations}

\remove{For the $N$-step reachable set $\mathcal{X}_N$, the number of non-zero elements in $G_N$ is constant, i.e., $\nnz{(G_N)}=\nnz{(G_s)} \; \forall N$, and $\nnz{(A_N)} \propto N$.}
Equations~\eqref{eq:reachability-trad-conzono} and \eqref{eq:reachability-alt-int-conzono} can be seen to be equivalent from the constrained zonotope equality constraints.

\setlength{\tabcolsep}{7pt}
\begin{table*}
    \centering
    \caption{\revmargin{1}{1.1}\protect\revres{1}{Memory complexity of the $N$-step reachable set $\mathcal{X}_N$ using the standard~\eqref{eq:reachability-trad}, graph of function~\eqref{eq:reachability-state-update}, and sparsity-promoting~\eqref{eq:reachability-alt-int} methods.}}
    \begin{tabular}{c|c c c c}
        \toprule
         Method & $n_G$ & $n_C$ & $\overline{\nnz}(G_N)$ & $\overline{\nnz}(A_N)$ \\ \midrule
         Eq.~\eqref{eq:reachability-trad} & $N(n_{G,s} + n_{G,u})$ & $N(n_{C,s} + n_{C,u} + n_x)$ & $N(n_x n_{G,u})$ & $N^2 (\frac{n_x  n_{G,u}}{2}) + N((\frac{n_x}{2} + n_{C,u})n_{G,u}$ \\
         & $+n_{G,0}$ & $+n_{C,0}$ & $+n_x  n_{G,0}$ & $+ (n_x + n_{C,s})n_{G,s} + n_x n_{G,0}) + n_{C,0} n_{G,0}$ \\ \midrule
         Eq.~\eqref{eq:reachability-state-update} & $2N(n_{G,s} + n_{G,u})$ & $N(2 n_{C,s} + n_{C,u} + 2 n_x + n_u)$ & $n_x (n_{G,s} + n_{G,u})$ & $N((3n_x + 2n_{C,s}) n_{G,s}$ \\
         & $+n_{G,0}$ & $+n_{C,0}$ & & $+(2 n_u + 2 n_{C,u} + 2n) n_{G,u}) + n_{C,0} n_{G,0}$ \\ \midrule
         Eq.~\eqref{eq:reachability-alt-int} & $N(n_{G,s} + n_{G,u})$ & $N(n_{C,s} + n_{C,u} + n_x)$ & $n_x  n_{G,s}$ & $N( (2n_x + n_{C,s}) n_{G,s} + (n_x + n_{C,u})n_{G,u})$ \\
         & $+n_{G,0}$ & $+n_{C,0}$ & & $+ n_{C,0} n_{G,0}$
    \end{tabular}
    \label{tab:memory-complexity}
\end{table*}

\subsubsection{Memory Complexity of the N-Step Reachable Set}
\revmargin{1}{1.1}\revres{1}{Table~\ref{tab:memory-complexity} gives the memory complexity for computing the set $\mathcal{X}_N$ with the standard, graph of function, and sparsity-promoting methods. The number of generators and constraints for sets $\mathcal{X}_0$, $\mathcal{S}$, and $\mathcal{U}$ are $(n_{G,0}, n_{C,0})$, $(n_{G,s}, n_{C,s})$, and $(n_{G,u}, n_{C,u})$, respectively. The number of states and inputs are $n_x$ and $n_u$. Upper bounds for the number of non-zero elements in a matrix are denoted as $\overline{\nnz}(\cdot)$. Bounds are computed by assuming that matrix products result in fully dense matrices. For clarity of exposition, we also assume that $G_{x,0}$, $A_{x,0}$, $G_u$, $A_u$, $G_s$, and $A_s$ are dense.}

\revres{1}{The standard and sparsity-promoting methods have the same number of generators and constraints, but the matrices $G_N$ and $A_N$ have much higher sparsity for the sparsity-promoting method. In particular, $\nnz{(G_N)} \propto 1$ and $\nnz{(A_N)} \propto N$ for~\eqref{eq:reachability-alt-int}, while $\nnz{(G_N)} \propto N$ and $\nnz{(A_N)} \propto N^2$ for~\eqref{eq:reachability-trad}.
To see that $\nnz{(A_N)} \propto N^2$ for~\eqref{eq:reachability-trad}, observe from~\eqref{eq:reachability-trad-conzono-A} that $\nnz{(A_N)} = \nnz{(A_{N-1})} + \nnz{(A G_{N-1})} + \cdots$, and therefore 
$\nnz{(A_N)} = \nnz{(A_0)} + \sum_{i=1}^N \nnz{(A G_{i-1})} + \cdots \propto \sum_{i=1}^N (i-1) \propto \frac{N(N-1)}{2}$.
When compared to the graph of function method, the sparsity-promoting method has fewer generators and constraints. The sparsity scaling is $\nnz{(G_N)} \propto 1$ and $\nnz{(A_N)} \propto N$ in both cases, but inspecting Table~\ref{tab:memory-complexity} shows that $\nnz{(G_N)}$ and $\nnz{(A_N)}$ are lower for~\eqref{eq:reachability-alt-int-conzono} than~\eqref{eq:reachability-state-update-conzono}.}

\subsection{Numerical Example: Reachable Sets of a Second-Order System}
Consider the discrete-time second-order linear system 
\begin{equation} \label{eq:second-order-sys}
    \begin{bmatrix}
        x_{k+1} \\ v_{k+1}
    \end{bmatrix} = \begin{bmatrix}
        1 & \Delta t \\
        -\omega_n^2 \Delta t & 1 - 2 \xi \omega_n \Delta t
    \end{bmatrix} \begin{bmatrix}
        x_k \\ v_k
    \end{bmatrix} + \begin{bmatrix}
        0 \\ \Delta t
    \end{bmatrix} u_k \;,
\end{equation}
with time step $\Delta t = 0.1$~s. The natural frequency and damping ratio are given as $\omega_n=0.3$~rad/s and $\xi=0.7$, respectively. The initial state set is $\mathcal{X}_0 = \left\{ \begin{bmatrix} x & v \end{bmatrix}^T \middle| x \in [-0.01, 0.01], v \in [0.49, 0.51] \right\}.$ The domain set of the states is $\mathcal{S} = \left\{ \begin{bmatrix} x & \dot{x} \end{bmatrix}^T \middle| x \in [-1,1], v \in [-1,1] \right\}$, and the domain set of the inputs is $\mathcal{U} = \left\{ u \middle| u \in [-1,1] \right\}$. All three of these sets are represented as zonotopes. 

Fig.~\ref{fig:reach-reach-sets} shows the reachable sets $\mathcal{X}_k$ for this system for $k \in \{0, ..., N\}$ where $N=15$. The sparsity patterns for the resulting constrained zonotope generator and constraint matrices $G_N$ and $A_N$ used to represent $\mathcal{X}_N$ are shown in Fig.~\ref{fig:reach-sparse}. The number of non-zero elements in $G_N$ and $A_N$ are given explicitly in Table~\ref{tab:nonzeros}. Reachability calculations using Eq.~\eqref{eq:reachability-alt-int} result in the constrained zonotope with the fewest non-zero elements in both $G_N$ and $A_N$. Further, this approach is tied with \eqref{eq:reachability-trad} for producing a constrained zonotope with the fewest number of generators $n_G$ and constraints $n_C$.

\begin{figure}[t]
    \centering
    \input{figs/reachability_reach_sets.pgf}
    \caption{Reachable sets $\mathcal{X}_k$ for the second-order linear system \eqref{eq:second-order-sys}. The gray set is the $N=15$-step reachable set $\mathcal{X}_{N}$.}
    \label{fig:reach-reach-sets}
\end{figure}

\begin{figure*}[t]
    \centering
    \input{figs/reachability_sparsity.pgf}
    \caption{Sparsity plots for matrices $G_N$ and $A_N$ where $\mathcal{X}_N = \left\langle G_N, \mathbf{c}_N, A_N, \mathbf{b}_N \right\rangle$ \revmargin{5}{5.6/5.7}\protect\revres{5}{using the standard~\eqref{eq:reachability-trad}, graph of function~\eqref{eq:reachability-state-update}, and sparsity-promoting~\eqref{eq:reachability-alt-int} reachability calculations to build the $N$-step reachable set $\mathcal{X}_N$.}}
    \label{fig:reach-sparse}
\end{figure*}

\begin{table}
    \caption{Number of non-zero elements in the generator matrix $G_N$ and constraint matrix $A_N$ for $\mathcal{X}_N$.}
    \centering
    \begin{tabular}{c|c c c}
        \toprule
         Method & Eq.~\eqref{eq:reachability-trad} & Eq.~\eqref{eq:reachability-state-update} & Eq.~\eqref{eq:reachability-alt-int} \\
         \midrule
         $\nnz{(G_N)}$ & 33 & 5 & 2 \\
         $\nnz{(A_N)}$ & 315 & 237 & 105
    \end{tabular}
    \label{tab:nonzeros}
\end{table}

\section{Alternating Direction Method of Multipliers for Constrained Zonotopes} \label{sec:ADMM}

\subsection{ADMM Formulation} \label{sec:admm-formulation}
We consider the problem of convex quadratic optimization over a constrained zonotope, i.e.,
\begin{equation} \label{eq:qp-over-conzono}
    \min_{\mathbf{x}} \frac{1}{2} \mathbf{x}^T P \mathbf{x} + \mathbf{q}^T \mathbf{x}, \; \text{s.t.} \; \mathbf{x} \in \mathcal{Z} \;,
\end{equation}
with $\mathcal{Z}$ a constrained zonotope and $P$ positive semi-definite. Referencing \eqref{eq:conzono-definition}, this can be re-written in terms of the constrained zonotope factors $\bm{\xi}$ as
\begin{subequations} \label{eq:qp-conzono-expanded}
\begin{align}
    &\min_{\bm{\xi}} \frac{1}{2} \bm{\xi}^T G^T P G \bm{\xi} + \left(G^T (P \mathbf{c} + \mathbf{q})\right)^T \bm{\xi} \;, \\
    &\text{s.t.} \; A \bm{\xi} = \mathbf{b},\; \bm{\xi} \in [-1,1]^{n_G}\;,
\end{align}
\end{subequations}
where constant terms have been neglected. Defining 
\begin{equation} \label{eq:tildeP-tildeq-def}
    \tilde{P} = G^T P G \;,\; \tilde{\mathbf{q}} = G^T (P \mathbf{c} + \mathbf{q}) \;,
\end{equation}
it is clear that \eqref{eq:qp-conzono-expanded} is a QP with constraints $\bm{\xi} \in \mathcal{A} \cap \mathcal{B}$ where $\mathcal{A} = \left\{ \bm{\xi} \in \real^{n_G} \middle| A \bm{\xi}=\mathbf{b} \right\}$ is an affine set and $\mathcal{B} = \left\{\bm{\xi} \in \real^{n_G} \middle| -\mathbf{1} \leq \bm{\xi} \leq \mathbf{1} \right\}$ is a box.

Problem~\eqref{eq:qp-conzono-expanded} can be written in the form of \eqref{eq:admm-general} for solution via ADMM by introducing optimization variables $\bm{\zeta}$. Using indicator functions, \eqref{eq:qp-conzono-expanded} then becomes
\begin{subequations} \label{eq:qp-admm-problem}
\begin{align}
&\min_{\bm{\xi}, \bm{\zeta}} \frac{1}{2} \bm{\xi}^T \tilde{P} \bm{\xi} + \tilde{\mathbf{q}}^T \bm{\xi} + I_{\mathcal{A}}(\bm{\xi}) + I_{\mathcal{B}}(\bm{\zeta}) \;, \\
&\text{s.t.}\;\bm{\xi} = \bm{\zeta} \;.
\end{align}
\end{subequations}

\revmargin{6}{6.4}\revres{6}{The ADMM iterations~\eqref{eq:admm-iterations-general} are then
\begin{subequations} \label{eq:admm-conzono-intermediate}
\begin{align}
    \bm{\xi}_{k+1} =& \argmin_{\bm{\xi}} \left( \frac{1}{2} \bm{\xi}^T (\tilde{P} + \rho I) \bm{\xi} + (\tilde{\mathbf{q}} + \rho (\mathbf{u}_k - \bm{\zeta}_k) )^T \bm{\xi} \right) \;,
    \nonumber \\
    &\text{s.t.}~A \bm{\xi} = \mathbf{b} \;, \label{eq:admm-conzono-intermediate-a} \\
    \bm{\zeta}_{k+1} =& \argmin_{\bm{\zeta}} ||\bm{\xi}_{k+1} - \bm{\zeta} + \mathbf{u}_k||^2_2 \;,\nonumber \\
    &\text{s.t.}~-\mathbf{1} \leq\bm{\zeta} \leq \mathbf{1} \;, \label{eq:admm-conzono-intermediate-b} \\
    \mathbf{u}_{k+1} =& \mathbf{u}_k + \bm{\xi}_{k+1} - \bm{\zeta}_{k+1} \;. \label{eq:admm-conzono-intermediate-c}
\end{align}   
\end{subequations}
Eq.~\eqref{eq:admm-conzono-intermediate-a} is an equality-constrained quadratic optimization problem and \eqref{eq:admm-conzono-intermediate-b} is a projection onto a box, so the iterations reduce to}
\begin{subequations} \label{eq:admm-implementation}
\begin{align}
&\bm{\xi}_{k+1} = \begin{bmatrix} I & 0 \end{bmatrix} M^{-1} \begin{bmatrix}
    -\tilde{\mathbf{q}} + \rho (\bm{\zeta}_k - \mathbf{u}_k) \\
    \mathbf{b}
\end{bmatrix} \;, \label{eq:admm-implementation-xi} \\
&\bm{\zeta}_{k+1} = \min(\max(\bm{\xi}_{k+1} + \mathbf{u}_k, \mathbf{-1}), \mathbf{1}) \;, \label{eq:admm-implementation-zeta} \\
&\mathbf{u}_{k+1} = \mathbf{u}_{k} + \bm{\xi}_{k+1} - \bm{\zeta}_{k+1} \label{eq:admm-implementation-u} \;,
\end{align}
\end{subequations}
where $\min(\cdot, \cdot)$ and $\max(\cdot, \cdot)$ denote element-wise operations, and 
\begin{equation} \label{eq:M-def}
    M = \begin{bmatrix}
    \tilde{P} + \rho I & A^T \\
    A & 0
\end{bmatrix} \;.
\end{equation}

For $M$ to be invertible, the following assumption must hold.
\begin{assumption} \label{as:A-full-row-rank}
    $\rank{(A)} = n_C$ \;.
\end{assumption}
If Assumption~\ref{as:A-full-row-rank} does not hold, then redundant constraints can be removed as in \cite[Alg. 1]{vinod2025projection} \revmargin{3}{3.2}\revres{3}{or using QR decomposition as in~\cite{bird2022hybrid}}.

To efficiently implement ADMM using \eqref{eq:admm-implementation}, $M$ is factorized once using a sparse LDLT factorization. Computing this factorization is typically the most expensive operation in the ADMM optimization and can be a significant fraction of the total solve time. The efficiency of computing this factorization and performing the subsequent back-solves in \eqref{eq:admm-implementation-xi} depends strongly on the number of non-zero elements in $M$.   

The ADMM formulation in \eqref{eq:admm-implementation} exploits the structure of the constrained zonotope set representation. As an illustrative example, consider the case where $\mathcal{Z}$ is a centrally symmetric hexagon $\mathcal{Z} = \left\langle G=\begin{bmatrix} \mathbf{g}_1 & \mathbf{g}_2 & \mathbf{g}_3 \end{bmatrix}, \mathbf{c}, [], [] \right\rangle$ and $P=I$. The matrix $M$ then becomes $M=G^TG + \rho I$, which has up to $9$ non-zero elements. 
\revmargin{3}{3.2}\revres{3}{In the sparsity-promoting iterations presented in Sec.~\ref{sec:sparsity}, much of the set complexity is moved to the constraint matrix $A$, so we additionally consider the representation $\mathcal{Z}=\left\langle \begin{bmatrix} I & 0 \end{bmatrix}, \mathbf{0}, \begin{bmatrix} I & -G \end{bmatrix}, \mathbf{0} \right\rangle$. In this case, $M$ will have up to 21 non-zero elements.}
Representing the same set using H-rep polytopes requires $6$ inequality constraints $\mathbf{a}_i^T \mathbf{x} \leq b_i$. The addition of $6$ slack variables $s_i>0$ transforms the inequality constraints into equality constraints such that the problem can be solved via a more general version of the ADMM formulation given in  \eqref{eq:admm-implementation} (see \cite{raghunathan2014admm, raghunathan2014infeasibility, raghunathan2014optimal}). The resulting $A$ matrix has up to $18$ non-zero elements in this case and $P+\rho I$ has $8$ non-zero elements due to the addition of the slack variables, so $M$ will have up to $44$ non-zero elements using H-rep.

Note that $M$ will not have fewer non-zero elements using a constrained zonotope representation $\mathcal{Z}$ vs. H-rep in general. In practice, $M$ has fewer non-zero elements when 1) sparsity-promoting operations are used to compute $\mathcal{Z}$ (Sec.~\ref{sec:sparsity}), and 2) the number of equality constraints in $\mathcal{Z}$ is minimized (e.g., by using operations on zonotopes to build $\mathcal{Z}$ when possible).

The ADMM formulation in \eqref{eq:admm-implementation} is more concise for optimizing over constrained zonotopes than other ADMM formulations. For example, the widely used ADMM solver OSQP \cite{stellato2020osqp} solves QPs with constraints of the form $C\mathbf{x} = \mathbf{z}$, $\mathbf{z}_l \leq \mathbf{z} \leq \mathbf{z}_u$. To optimize over $\mathcal{Z}$ using this form requires $C = \begin{bmatrix} A^T & I^T \end{bmatrix}^T$\remove{, which}\revmargin{5}{5.1}\revres{5}{. This} adds $2n_G$ non-zero elements to the linear system that OSQP must solve and adds $n_C$ projection operations.

Combining constrained zonotopes with high sparsity and the ADMM iterations \eqref{eq:admm-implementation} results in an approach to solving QPs that scales well to large problem sizes\revmargin{5}{5.3}\remove{, as will be shown in Sec.~\ref{sec:results-mpc}}.

\subsection{Infeasibility Detection}
Eq.~\eqref{eq:qp-admm-problem} is infeasible if there do not exist $\bm{\xi} \in \mathcal{A}$ and $\bm{\zeta} \in \mathcal{B}$ such that $\bm{\xi} = \bm{\zeta}$. In \cite{raghunathan2014infeasibility}, infeasibility detection was studied for ADMM problems of a form identical to \eqref{eq:qp-admm-problem}, except $Z^T \tilde{P} Z$ is required to be positive definite where $Z$ is a basis for the null space of $A$. The authors propose practical infeasibility detection criteria based on several conditions involving the ADMM iterates being satisfied within some user-defined tolerances.

In the following theorem, we propose an infeasibility certificate for \eqref{eq:qp-admm-problem} that is exact within machine precision and does not require any user-defined tolerances. 
\revmargin{1}{1.2}\revres{1}{The interval notation $[\bm{\beta}]$ is used to denote the set $\mathcal{B} = \left\{\bm{\xi} \in \real^{n_G} \middle| -\mathbf{1} \leq \bm{\xi} \leq \mathbf{1} \right\}$.}

\begin{theorem} \label{thm:inf-certificate}
    \revmargin{3}{3.3}
     Problem~\eqref{eq:qp-admm-problem} is infeasible if $\exists \mathbf{v} \in \text{span}(A^T)$ and \revres{3}{$\exists \tilde{\bm{\xi}} \in \mathcal{A}$}
     such that $\mathbf{v}^T \tilde{\bm{\xi}} \notin \mathbf{v}^T [\bm{\beta}]$.
     \revmargin{6}{6.5}
    
    \begin{proof}
        The vector $\mathbf{v}$ can be written as $\mathbf{v} = A^T \bm{\lambda}, \; \bm{\lambda} \in \real^{n_C}$. For all $\bm{\xi} \in \mathcal{A}$, $A \bm{\xi} = \mathbf{b}$, and therefore $\mathbf{v}^T \bm{\xi} = \bm{\lambda}^T \mathbf{b}$ $\forall \bm{\xi} \in \mathcal{A}$ and $\min_{\bm{\xi} \in \mathcal{A}}( \mathbf{v}^T \bm{\xi}) = \max_{\bm{\xi} \in \mathcal{A}} (\mathbf{v}^T \bm{\xi}) = \mathbf{v}^T \tilde{\bm{\xi}}$. 
        Defining $[c] = \mathbf{v}^T [\bm{\beta}] = [\underline{c}, \overline{c}]$, $\min_{\bm{\xi} \in \mathcal{B}}( \mathbf{v}^T \bm{\xi}) = \underline{c}$ and $\max_{\bm{\xi} \in \mathcal{B}}( \mathbf{v}^T \bm{\xi}) = \overline{c}$. If $\mathbf{v}^T \tilde{\bm{\xi}} \notin [c]$, then $\min_{\bm{\xi} \in \mathcal{B}}( \mathbf{v}^T \bm{\xi}) > \max_{\bm{\xi} \in \mathcal{A}} (\mathbf{v}^T \bm{\xi})$ or $\min_{\bm{\xi} \in \mathcal{A}}( \mathbf{v}^T \bm{\xi}) > \max_{\bm{\xi} \in \mathcal{B}} (\mathbf{v}^T \bm{\xi})$, and by the separating hyperplane theorem $\mathcal{A} \cap \mathcal{B} = \emptyset$ \cite{rockafellar1997convex}, so \eqref{eq:qp-admm-problem} is infeasible.
    \end{proof}
\end{theorem}

The following corollary guarantees that if \eqref{eq:qp-admm-problem} is infeasible and the condition $Z^T \tilde{P} Z > 0$ holds, then an infeasibility certificate of the form given in Thm.~\ref{thm:inf-certificate} is generated 
for $k$ sufficiently large.

\begin{corollary} \label{cor:inf-conv}
    If \eqref{eq:qp-admm-problem} is infeasible and $Z^T \tilde{P} Z$ is positive definite where $Z$ is a basis for the null space of $A$, then the difference of the iterates \revmargin{6}{6.6}$\bm{\zeta}_k - \bm{\xi}_k \revres{6}{=} \mathbf{v}$ for $k \geq k'$ with $k'$ sufficiently large, where $\mathbf{v}$ is an infeasibility certificate for problem~\eqref{eq:qp-admm-problem}.

    \begin{proof}
    From \cite[Thm. 1]{raghunathan2014infeasibility}, if \eqref{eq:qp-admm-problem} is infeasible and $Z^T \tilde{P} Z>0$, then for $k>k'$, $\bm{\zeta}_k - \bm{\xi}_k = \bm{\lambda}^o \in \text{span}{(A^T)}$ where $\bm{\lambda}^o = \bm{\zeta}^o - \bm{\xi}^o$ with $\bm{\xi}^o$ and $\bm{\zeta}^o$ the solutions to $\argmin_{\bm{\xi}, \bm{\zeta}} ||\bm{\zeta} - \bm{\xi}||^2$ s.t. $\bm{\xi} \in \mathcal{A}$ and $\bm{\zeta} \in \mathcal{B}$.
    Since \eqref{eq:qp-admm-problem} is infeasible, $||\bm{\zeta}^o - \bm{\xi}^o||^2 = (\bm{\zeta}^o - \bm{\xi}^o)^T(\bm{\zeta}^o-\bm{\xi}^o) > 0$. Because $\bm{\xi}^o$ and $\bm{\zeta}^o$ minimize $||\bm{\zeta}^o - \bm{\xi}^o||^2$ for 
    \revmargin{3}{3.4}\revres{3}{$\begin{bmatrix}\bm{\xi}^T & \bm{\zeta}^T \end{bmatrix}^T \in \mathcal{A} \times \mathcal{B}$ and $\mathcal{A} \times \mathcal{B}$ is convex,} 
    \revres{1}{by the first order optimality criterion~\cite{boyd2004convex},}\revmargin{1}{1.3}
    $(\bm{\zeta}^o - \bm{\xi}^o)^T(\bm{\zeta}-\bm{\xi}) \geq (\bm{\zeta}^o - \bm{\xi}^o)^T(\bm{\zeta}^o-\bm{\xi}^o) > 0$ 
    \revres{3}{$\forall \begin{bmatrix}\bm{\xi}^T & \bm{\zeta}^T \end{bmatrix}^T \in \mathcal{A} \times \mathcal{B}$.}
    Thus, $(\bm{\lambda}^o)^T \bm{\xi} < (\bm{\lambda}^o)^T \bm{\zeta}$ $\forall \bm{\xi} \in \mathcal{A}$ and $\bm{\zeta} \in \mathcal{B}$ and by Thm.~\ref{thm:inf-certificate}, $\mathbf{v} = \bm{\lambda}^o$ constitutes an infeasibility certificate.
    \end{proof}
\end{corollary}

In some cases, one may be interested only in whether an infeasibility certificate for \eqref{eq:qp-admm-problem} exists. For example, when using constrained zonotopes to perform reachability analysis, it may be necessary to ensure $\mathcal{X} \cap \mathcal{O} = \emptyset$ where $\mathcal{X}$ is a reachable set of the system dynamics and $\mathcal{O}$ is an unsafe set. In this case, one can choose $\tilde{P} = I$ to ensure $Z^T \tilde{P} Z > 0$.

To implement Thm.~\ref{thm:inf-certificate} for infeasibility detection, the difference of the iterates $\bm{\zeta}_k - \bm{\xi}_k$ is projected onto the span of $A^T$ to generate a candidate infeasibility certificate vector $\mathbf{v}$. If $\mathbf{v}$ satisfies Thm.~\ref{thm:inf-certificate}, then \eqref{eq:qp-admm-problem} is infeasible regardless of whether $k \geq k'$ as defined in Cor.~\ref{cor:inf-conv}. In practice, this results in infeasibility certificates being generated in as few as $k=1$ iterations for many problems\revmargin[false]{5}{5.3}\remove{, as will be shown in Sec.~\ref{sec:results-reachability}}. Empirically, infeasibility certificates are produced using this implementation for $Z^T \tilde{P} Z$ positive semi-definite as well.

\subsection{Interpretation of ADMM Iterations}
The ADMM iterates alternately enforce the constrained zonotope equality constraints $A \bm{\xi} = \mathbf{b}$ and box constraints $\bm{\xi} \in [-1,1]^{n_G}$. For the case that the constrained zonotope $\mathcal{Z}$ is 
constructed by applying the set operations in \eqref{eq:set-ops} to zonotopes, the equality constraints in $\mathcal{Z}$ only enforce the intersection operations. All other constraints in the optimization problem are enforced via the box constraints. This decoupling of the constraints is useful in certain scenarios and may justify early termination of the ADMM iterations.

Consider an MPC optimization problem for the system \eqref{eq:linsys} for which the feasible set $\mathcal{Z}$ is defined such that any point $\mathbf{z} \in \mathcal{Z}$ satisfies 
\begin{subequations}
\begin{align}
    &\mathbf{z} = \begin{bmatrix}
             \mathbf{x}_0^T & \mathbf{u}_0^T & \cdots & \mathbf{u}_{N-1}^T & \mathbf{x}_N^T
        \end{bmatrix}^T \;, \label{eq:Z_def_a} \\
    &\mathbf{x}_0 \in \mathcal{X}_0 \;, \label{eq:Z_def_b} \\
    &\forall k \in \{0,..,N-1\}: \nonumber \\
    &\hphantom{\forall k} \mathbf{x}_k \in \mathcal{S},\; \mathbf{u}_k \in \mathcal{U},\; \mathbf{x}_{k+1} \in \{A\mathbf{x}_k + B\mathbf{u}_k\} \cap \mathcal{S} \;. \label{eq:Z_def_c}
\end{align}
\end{subequations}

The sets $\mathcal{X}_k$ can be built from the iterations \eqref{eq:reachability-trad}, \eqref{eq:reachability-state-update}, or \eqref{eq:reachability-alt-int} (see Algorithm~\ref{alg:mpc-reachability} for an implementation using a variant of \eqref{eq:reachability-alt-int}). If \eqref{eq:reachability-trad} is used and $\mathcal{X}_0$, $\mathcal{U}$, and $\mathcal{S}$ are all zonotopes, then the equality constraints in $\mathcal{Z}$ only enforce $\mathcal{X}_k \subseteq \mathcal{S}$. In this case, the $\bm{\zeta}_k$ iterates of \eqref{eq:admm-implementation} are feasible with respect to the system dynamics and input constraints but may not satisfy the state constraints $\mathbf{x}_k \in \mathcal{S}$. This separation is especially useful for the case that there are no state constraints, in which case the ADMM iterates $\bm{\zeta}_k$ are always primal feasible. Only one iteration is then required to produce a feasible solution to \eqref{eq:qp-over-conzono}, and subsequent iterations only serve to improve the optimality of the solution.

In \cite{srikanthan2024closed}, an MPC formulation using ADMM with early termination was presented that decoupled enforcement of system dynamics constraints from state and input feasibility constraints between ADMM steps. In that formulation, the equality constraints enforced in \eqref{eq:admm-gen-a} correspond to the system dynamics while feasibility constraints are enforced in \eqref{eq:admm-gen-b}. This is analogous to building the MPC problem using \eqref{eq:reachability-alt-int} with zonotopes in our framework. \remove{As described above,}\revmargin{5}{5.3} Using \eqref{eq:reachability-trad} with zonotopes to build the MPC problem would instead simultaneously enforce dynamics and input constraints with state constraints decoupled.

\subsection{ADMM Algorithm for Constrained Zonotopes}
ADMM is a first-order method, and as such the convergence rate is sensitive to problem conditioning \cite{boyd2004convex}. Pre-conditioning algorithms are frequently used to mitigate this challenge \cite{rey2016admm, chari2024constraint, stellato2020osqp}. 

One approach to improving the conditioning of a problem is to introduce a change of variables $\tilde{\mathbf{x}} = D \mathbf{x}$ with $D$ diagonal such that the optimal variables $\tilde{x}_i^*$ all have similar magnitude \cite{nocedal1999numerical}. By optimizing over the constrained zonotope factors $\bm{\xi}$, the optimization variables are well-scaled by construction (i.e., they are all normalized to $\xi_i \in [-1,1]$). As such, we do not use any pre-conditioning algorithm, thus simplifying the implementation.

The complete ADMM algorithm for solving \eqref{eq:qp-over-conzono} is given by Algorithm~\ref{alg:admm}. Algorithm~\ref{alg:admm} returns an $\mathbf{x}^*$ which approximately solves \eqref{eq:qp-over-conzono} if the convergence criteria are satisfied. In this algorithm, the constraint set is given as $\mathcal{Z} = \left\langle G, \mathbf{c}, A, \mathbf{b} \right\rangle$. 
The parameter $k_{\text{inf}}$ is the frequency at which the algorithm looks for an infeasibility certificate. 
The length of vectors $\bm{\xi}_k$, $\bm{\zeta}_k$, and $\mathbf{u}_k$ is $n$. In practice, Algorithm~\ref{alg:admm} may be modified to include a time limit or maximum number of iterations.
 
\begin{algorithm}
\caption{ADMM for solving \eqref{eq:qp-over-conzono} \\
\revmargin{1}{1.4}\protect\revres{1}{Inputs: $\mathcal{Z}$, $P$, $\mathbf{q}$, $\rho$, $\epsilon_p$, $\epsilon_d$} \\
\protect\revres{1}{Outputs: $\mathbf{x}^* \in \mathcal{Z}$, infeasibility certificate flag}
}
\begin{algorithmic}[1]
    \STATE factorize $M$ and $A A^T$
    \STATE $k \gets 0$
    \STATE $(\bm{\xi}_k, \bm{\zeta}_k, \mathbf{u}_k) \gets (\mathbf{0}, \mathbf{0}, \mathbf{0})$
    \STATE (converged, infeasible) $\gets$ (false, false)
    \WHILE{\textbf{not} converged \textbf{and not} infeasible}
        \STATE $\bm{\xi}_{k+1} = \begin{bmatrix} I & 0 \end{bmatrix}        M^{-1} \begin{bmatrix} -\tilde{\mathbf{q}} + \rho           (\bm{\zeta}_k - \mathbf{u}_k) \\
            \mathbf{b}
            \end{bmatrix}$
        \STATE $\bm{\zeta}_{k+1} = \min(\max(\bm{\xi}_{k+1} + \mathbf{u}_k, \mathbf{-1}), \mathbf{1})$
        \STATE $\mathbf{u}_{k+1} = \mathbf{u}_{k} + \bm{\xi}_{k+1} - \bm{\zeta}_{k+1}$ 
        \IF{$k \; \% \; k_{\text{inf}} = 0$}
            \STATE $\mathbf{v} \gets A^T (A A^T)^{-1} A (\bm{\zeta}_{k+1} - \bm{\xi}_{k+1})$
            \STATE infeasible $\gets (\mathbf{v}^T \bm{\xi}_{k+1} \notin \mathbf{v}^T [\bm{\beta}])$
        \ENDIF
        \STATE $\mathbf{r}_p \gets \bm{\xi}_{k+1} - \bm{\zeta}_{k+1}$
        \STATE $\mathbf{r}_d \gets \rho (\bm{\zeta}_{k+1} - \bm{\zeta}_k)$
        \STATE converged $\gets (||\mathbf{r}_p||_2 < \sqrt{n} \epsilon_p)$ \textbf{and} $(||\mathbf{r}_d||_2 < \sqrt{n} \epsilon_d)$
        \STATE $k \gets k+1$
    \ENDWHILE
    \STATE $\mathbf{x}^* \gets G \bm{\zeta}_{k+1} + \mathbf{c}$
    \RETURN $\mathbf{x}^*$, infeasible
\end{algorithmic}
\label{alg:admm}
\end{algorithm}
\section{Applications} \label{sec:applications}
The following application examples demonstrate the utility of combining the sparse reachability operations presented in Sec.~\ref{sec:sparsity} with the ADMM algorithm presented in Sec.~\ref{sec:ADMM} to solve control, estimation, and safety verification problems. 
\revmargin[true]{6}{6.7}\revres{6}{In all of the presented examples, efficient problem formulations rely on the sparsity-promoting iterations~\eqref{eq:reachability-alt-int}.}

\subsection{Model Predictive Control} \label{sec:applications-mpc}

Consider an MPC problem with time-varying state constraints
\begin{subequations} \label{eq:results-mpc-prob}
\begin{align}
    \min_{\mathbf{x}, \mathbf{u}} &\sum_{k=0}^{N-1} \left( (\mathbf{x}_k-\mathbf{x}_k^r)^T Q (\mathbf{x}_k-\mathbf{x}_k^r) + \mathbf{u}_k^T R \mathbf{u}_k \right) + \nonumber \\ 
    &\qquad\qquad\;\; (\mathbf{x}_N - \mathbf{x}_N^r)^T Q_N (\mathbf{x}_N - \mathbf{x}_N^r) \;, \\
    \text{s.t. } &\forall k \in \{0, ..., N-1\}: \nonumber \\
    &\mathbf{x}_{k+1} = A \mathbf{x}_k + B \mathbf{u}_k \;,\; \mathbf{x}_{k=0} = \mathbf{x}_0 \;, \\
    &\revres{4}{\mathbf{x}_k \in \mathcal{S}_k \;,\; \mathbf{x}_N \in \mathcal{S}_N}\revmargin{4}{4.1} \;,\; \mathbf{u}_k \in \mathcal{U} \;, 
\end{align}
\end{subequations}
where $\mathbf{x}_k$ are the system states, $\mathbf{u}_k$ are the control inputs, $\mathbf{x}_k^r$ are reference states, and $\mathbf{x}_0$ is the initial condition. The input constraint set is $\mathcal{U}$ and $\mathcal{S}_k$ is the time-varying state constraint set. The MPC horizon is $N$. The state and input stage cost matrices are $Q$ and $R$, and the terminal cost matrix is $Q_N$.

If the sets $\mathcal{S}_k$ and $\mathcal{U}$ are zonotopes or constrained zonotopes, then this MPC problem can be put into the form of \eqref{eq:qp-over-conzono} using Algorithm~\ref{alg:mpc-reachability}. Note that line~\ref{alg-line:mpc-sparse-reach} in Algorithm~\ref{alg:mpc-reachability} implements the sparse reachable set identity given in \eqref{eq:reachability-alt-int}.

\begin{algorithm}
    \caption{Build MPC optimization problem in form of \eqref{eq:qp-over-conzono} using reachability analysis of constrained zonotopes \\
    \revmargin{1}{1.4}\protect\revres{1}{Inputs: $\mathbf{x}_0$, $\mathbf{x}^r_k$, $Q$, $R$, $Q_N$, $A$, $B$, $N$} \\
    \protect\revres{1}{Outputs: $\mathcal{Z}, P, \mathbf{q}$}}
    \begin{algorithmic}[1]
        \STATE $\mathcal{Z} \gets \{\mathbf{x}_0\}$
        \STATE $P \gets Q$
        \STATE $\mathbf{q} \gets \mathbf{0}$ 
        \FOR{$k \in \{1, ..., N\}$}
            \STATE $\mathcal{Z} \gets (\mathcal{Z} \times \mathcal{U} \times \mathcal{S}_{k}) \cap_{\begin{bmatrix}
                0 & \dots & 0 & A & B & -I
            \end{bmatrix}} \{\mathbf{0}\}$ \label{alg-line:mpc-sparse-reach}
            \IF{$k = N$}
                \STATE $P \gets \text{blkdiag}\left( \begin{bmatrix} P & R & Q_N \end{bmatrix} \right)$
            \ELSE{}
                \STATE $P \gets \text{blkdiag}\left( \begin{bmatrix} P & R & Q  \end{bmatrix} \right)$
            \ENDIF
            
            \STATE $\mathbf{q} \gets \begin{bmatrix} \mathbf{q}^T & \mathbf{0}^T & -(Q\mathbf{x}^r_k)^T \end{bmatrix}^T$
        \ENDFOR
        \RETURN $(\mathcal{Z}, P, \mathbf{q})$
    \end{algorithmic}
    \label{alg:mpc-reachability}
\end{algorithm}

\subsection{Moving Horizon Estimation and Set-Valued State Estimation} \label{sec:app-mhe}

Consider the problem of estimating the state $\mathbf{x}_k$ from measurements $\mathbf{y}_k$ of the linear system
\begin{subequations} \label{eq:mhe-system}
\begin{align}
    &\mathbf{x}_{k+1} = A \mathbf{x}_k + B \mathbf{u}_k + \mathbf{w}_k \;,\; \mathbf{y}_k = C\mathbf{x}_k + \mathbf{v}_k \;, \\
    &\text{s.t.} \;\forall k: \mathbf{x}_k, \mathbf{x}_{k+1} \in \mathcal{S} \;,\; \mathbf{w}_k \in \mathcal{W} \;,\; \mathbf{v}_k \in \mathcal{V} \;,
\end{align}
\end{subequations}
subject to bounded process and measurement noise $\mathbf{w}_k$ and $\mathbf{v}_k$, respectively. The state domain set $\mathcal{S}$ is discussed in Sec.~\ref{sec:sparsity}.
In the context of SVSE, constrained zonotopes have been used to find a bounding set $\mathcal{X}_k$ such that $\mathbf{x}_k \in \mathcal{X}_k$ \cite{scott2016constrained}. 
While SVSE algorithms produce a set that is guaranteed to contain the state of the system, they do not in general produce a best state estimate within that set. The following presents an MHE implementation based on reachability analysis with constrained zonotopes that produces both a bounding set of system states and a best estimate of the state within that set.

Assuming $\mathcal{S}$, $\mathcal{W}$, $\mathcal{V}$, and $\mathcal{X}_{-N}$ are zonotopes or constrained zonotopes, reachability analysis with constrained zonotopes can be used to solve the SVSE problem for \eqref{eq:mhe-system} via the iterations
\begin{equation} \label{eq:set-valued-state-est-scott}
    \mathcal{X}_{k+1} = \Big( (A \mathcal{X}_k \oplus B \mathbf{u}_k \oplus \mathcal{W}) \cap_C (\mathbf{y}_{k+1} \oplus (-\mathcal{V})) \Big) \cap \mathcal{S} \;,
\end{equation}
where $\mathcal{X}_k$ is the set of possible states of the system at time step $k$~\cite{scott2016constrained}. Eq.~\eqref{eq:set-valued-state-est-scott} can be equivalently written in a high-sparsity form by modifying \eqref{eq:reachability-alt-int}, giving
\begin{multline} \label{eq:set-valued-state-est-sparse}
    \mathcal{X}_{k+1} = \begin{bmatrix} 0 & 0 & I \end{bmatrix} 
    \Big((\mathcal{X}_{k} \times \mathcal{W} \times (\mathcal{S} \cap_C (\mathbf{y}_{k+1} \oplus (-\mathcal{V})))) \\ \cap_{\begin{bmatrix} A & I & -I \end{bmatrix}} \{-B \mathbf{u}_{k} \} \Big) \;.
\end{multline}

Next, consider the MHE problem formulation for the system given in \eqref{eq:mhe-system}
\begin{subequations} \label{eq:mhe}
\begin{align}
    \min_{\mathbf{w}, \mathbf{v}, \mathbf{x}_{-N}} &\sum_{k=-N}^{-1} \left[ \mathbf{w}_k^T Q^{-1} \mathbf{w}_k + \mathbf{v}_{k+1}^T R^{-1} \mathbf{v}_{k+1} \right] + \nonumber \\
    &\qquad (\mathbf{x}_{-N}-\hat{\mathbf{x}}_{-N})^T P^{-1}_{-N} (\mathbf{x}_{-N}-\hat{\mathbf{x}}_{-N}) \;, \\
    \text{s.t.} \; &\forall k \in \{-N, ..., -1 \} : \nonumber \\
    &\mathbf{x}_{k+1} = A \mathbf{x}_k + B \mathbf{u}_k + \mathbf{w}_k \;,\; \mathbf{y}_k = C\mathbf{x}_k + \mathbf{v}_k \;, \\
    &\mathbf{x}_k, \mathbf{x}_0 \in \mathcal{S} \;,\; \mathbf{w}_k \in \mathcal{W} \;,\; \mathbf{v}_k \in \mathcal{V} \;,\; \mathbf{x}_{-N} \in \mathcal{X}_{-N} \;.
\end{align}
\end{subequations}
The process and measurement noise covariance matrices are $Q$ and $R$, respectively. The covariance of the state estimate at time step $k=-N$ is $P_{-N}$\remove{, and}\revmargin{5}{5.1}\revres{5}{. T}he prior state estimate for $k=-N$ is $\hat{\mathbf{x}}_{-N}$. The set of possible states at time step $k=-N$ is $\mathcal{X}_{-N}$, which is updated recursively between MHE iterations by applying \eqref{eq:set-valued-state-est-sparse}.

Algorithm~\ref{alg:mhe} shows how the MHE optimization problem \eqref{eq:mhe} can be written in the form of \eqref{eq:qp-over-conzono} using constrained zonotopes and the set operations in \eqref{eq:set-ops}. Line~\ref{alg-line:mhe-reach} implements \eqref{eq:set-valued-state-est-sparse} without projecting onto $\mathcal{X}_{k+1}$. 
Elements $\mathbf{z} \in \mathcal{Z}$ where $\mathcal{Z}$ is the set of feasible solutions to~\eqref{eq:set-valued-state-est-sparse} have the structure
\begin{equation}
    \mathbf{z} = \begin{bmatrix} \mathbf{x}_{-N}^T & \mathbf{w}_{-N}^T & \cdots & \mathbf{w}_{-1}^T & \mathbf{x}_0^T \end{bmatrix} \;.
\end{equation}

\remove{and }\revmargin{5}{5.1}\revres{5}{A}s such, the solution $\mathcal{X}_0$ to the SVSE problem at time step $k=0$ can be readily extracted via a linear map operation. After solving \eqref{eq:qp-over-conzono}, a best state estimate $\hat{\mathbf{x}}_0 \in \mathcal{X}_0$ can also be extracted.

\begin{algorithm}
\caption{Build MHE problem in form of \eqref{eq:qp-over-conzono} and return set of possible states at time step $k=0$. \\
\revmargin{1}{1.4}\protect\revres{1}{Inputs: $\mathcal{X}_{-N}$, $\mathcal{W}$, $\mathcal{V}$, $\hat{\mathbf{x}}_{-N}$, $\mathbf{y}_k$, $\mathbf{u}_k$, $P^{-1}_{-N}$, $A$, $B$, $C$, $Q$, $R$, $N$} \\
\protect\revres{1}{Outputs: $\mathcal{Z}$, $P$, $\mathbf{q}$, $\mathcal{X}_0$}
}
\begin{algorithmic}[1]
    \STATE $\mathcal{Z} \gets \mathcal{X}_{-N}$ 
    \STATE $P \gets P^{-1}_{-N}$
    \STATE $\mathbf{q} = -P_{-N}^{-1} \hat{\mathbf{x}}_{-N}$
    \FOR{$k \in \{-N, ..., -1\}$}
        \STATE $\begin{aligned} \mathcal{Z} \gets &\left( \mathcal{Z} \times \mathcal{W} \times (\mathcal{S} \cap_C (\mathbf{y}_{k+1} \oplus (-\mathcal{V}))) \right) \\ &\cap_{\begin{bmatrix} 0 & \dots & 0 & A & I & -I\end{bmatrix}} \{-B \mathbf{u}_k\} \end{aligned}$
        \label{alg-line:mhe-reach}
        \STATE $P \gets \text{blkdiag}\left( \begin{bmatrix} P & Q^{-1} & C^T R^{-1} C \end{bmatrix} \right)$
        \STATE $\mathbf{q} \gets \begin{bmatrix} \mathbf{q}^T & \mathbf{0}^T & -(C^T R^{-1} \mathbf{y}_{k+1})^T \end{bmatrix}^T$
    \ENDFOR
    \STATE $\mathcal{X}_0 \gets \begin{bmatrix} 0 & \dots & 0 & I \end{bmatrix} \mathcal{Z}$
    \RETURN $\left( \mathcal{Z}, P, \mathbf{q} \right)$, $\mathcal{X}_0$
\end{algorithmic}
\label{alg:mhe}
\end{algorithm}

\subsection{Safety Verification} \label{sec:app-safety}
The infeasibility certificate given in Thm.~\ref{thm:inf-certificate} can be used for safety verification. Consider a disturbed linear system
\begin{subequations}
\begin{align}
    &\mathbf{x}_{k+1} = A \mathbf{x}_k + B \mathbf{u}_k + \mathbf{w}_k \;, \\
    &\text{s.t.} \; \forall k: \mathbf{x}_k, \mathbf{x}_{k+1} \in \mathcal{S} \;,\; \mathbf{w}_k \in \mathcal{W} \;,
\end{align}
\end{subequations}
subject to the state feedback control law $\mathbf{u}_k = -K(\mathbf{x}_k-\mathbf{x}_k^r)$ where $\mathbf{x}_k^r$ is a reference state. The state domain set $\mathcal{S}$ is discussed in Sec.~\ref{sec:sparsity}. Neglecting any input saturation, the reachable sets of the system states are defined by the sparsity-promoting iterations
\begin{equation}
    \mathcal{X}_{k+1} = \left( \begin{bmatrix} 0 & 0 & I \end{bmatrix} 
    (\mathcal{X}_{k} \times \mathcal{W} \times \mathcal{S}) \cap_{[A_c \; I \; -I ]} \{-B \mathbf{u}_{\mathit{ff}} \} \right) \;,
\end{equation}
where $A_c = A - BK$ and $\mathbf{u}_{\mathit{ff}} = K \mathbf{x}_k^r$. Given some unsafe set $\mathcal{O}$, possibly in a lower-dimensional space than the state vector, trajectories of the system are safe at time step $k$ if $R\mathbf{x} \notin \mathcal{O} \; \forall \mathbf{x} \in \mathcal{X}_k$, where $R$ maps $\mathbf{x}$ onto the safety-related subspace. This condition can be verified if Algorithm~\ref{alg:admm} finds an infeasibility certificate for \eqref{eq:qp-admm-problem}, with $\mathcal{A}$ and $\mathcal{B}$ the equality and inequality constraints of $\mathcal{Z} = \mathcal{X}_k \cap_R \mathcal{O}$. 

\section{Numerical Results} \label{sec:numerical-examples}

\subsection{Implementation}

The following numerical examples apply the methods of Sec.~\ref{sec:applications} in the context of motion control, localization, and online safety verification problems for autonomous mobile robots. In all cases, the discrete-time double integrator model 
\begin{subequations} \label{eq:results-dbl-int-dyn}
\begin{align}
    &\mathbf{x}_k = \begin{bmatrix} x_k & y_k & v_{x,k} & v_{y,k} \end{bmatrix}^T \;, \\
    &\mathbf{u}_k = \begin{bmatrix} a_{x,k} & a_{y,k} \end{bmatrix}^T \;, \\
    &A = \begin{bmatrix}
        1 & 0 & \Delta t & 0 \\
        0 & 1 & 0 & \Delta t \\
        0 & 0 & 1 & 0 \\
        0 & 0 & 0 & 1
    \end{bmatrix} \;,\; B = \begin{bmatrix}
        \frac{\Delta t^2}{2} & 0 \\
        0 & \frac{\Delta t^2}{2} \\
        \Delta t & 0 \\
        0 & \Delta t
    \end{bmatrix} \;,
\end{align}
\end{subequations}
is used, where $x_k$ and $y_k$ are position states, $v_{x,k}$ and $v_{y,k}$ are velocity states, and $a_{x,k}$ and $a_{y,k}$ are accelerations.
Using differential flatness, the double integrator model can be shown to be equivalent to the nonlinear unicycle model, which is widely used to model autonomous mobile robots~\cite{diffflat}.

All examples are implemented using the open-source toolbox ZonoOpt
, which was developed as a contribution of this paper. This toolbox provides classes for zonotopes, constrained zonotopes, and hybrid zonotopes, implements the set operations in \eqref{eq:set-ops} (among others), and includes an implementation of Algorithm~\ref{alg:admm}. ZonoOpt is implemented in C++ and Python bindings are available for all classes and functions. Sparse linear algebra is utilized throughout using the Eigen linear algebra library \cite{eigenweb}. ZonoOpt is influenced by the MATLAB toolbox zonoLAB \cite{koeln2023zonolab}.

Unless otherwise stated, the following examples use ADMM penalty parameter $\rho=1$ and primal and dual convergence tolerances of $\epsilon_p=\epsilon_d=0.01$. The infeasibility checking interval is $k_{\text{inf}} = 10$.

\subsection{Model Predictive Control} \label{sec:results-mpc}

\begin{figure}[t]
    \centering
    \input{figs/mpc_time_varying_cons.pgf}
    \caption{Optimal trajectory for the MPC optimization problem with $N=55$. The blue sets are the time-varying position constraint sets $\mathcal{P}_k$, and the black line corresponds to the reference states $\mathbf{x}^r_k$. The red dots are the optimized trajectory produced by the ADMM solver in ZonoOpt. The green obstacle polytopes are displayed only for illustrative purposes and are not directly considered in the MPC problem formulation.}
    \label{fig:mpc-traj}
\end{figure}

Consider the MPC problem described in Sec.~\ref{sec:applications-mpc} using the discrete-time double integrator model~\eqref{eq:results-dbl-int-dyn}.
\revmargin[true]{5}{5.4}\revres{5}{This problem is depicted in Fig.~\ref{fig:mpc-traj}.}
The time-varying constraint set $\mathcal{S}_k$ is decomposed into a time-varying position component and a constant velocity component as $\mathcal{S}_k = \mathcal{P}_k \times \mathcal{V}$. The velocity and input constraints $\mathcal{V}$ and $\mathcal{U}$ are regular 12-sided polygons, and time-varying position constraints are denoted by $\mathcal{P}_k$. In this example, the $\mathcal{P}_k$ are given by centrally symmetric hexagons. 
\revmargin{6}{6.8}\revres{6}{The $\mathcal{P}_k$ define a corridor that the system must remain inside, and the position references $\mathbf{x}_k^r$ are centered within this corridor.
For motion planning problems, the $\mathcal{P}_k$ may be specified by a higher-level autonomous planning algorithm such as in \cite{zhang2023hierarchical}.}
The regular polygon constraints approximate the quadratic constraints $v_{x,k}^2 + v_{y,k}^2 \leq v_{\text{max}}^2$ and $a_{x,k}^2 + a_{y,k}^2 \leq (v_{\text{min}}\omega_{\text{max}})^2$, where $v_{\text{min}}$,  $v_{\text{max}}$, and $\omega_{\text{max}}$ are linear and angular velocity limits \cite{whitaker2021optimal}.

In this example, the state vector has dimensions of $\mathbf{x} \sim \begin{bmatrix} \text{m} & \text{m} & \text{m/s} & \text{m/s} \end{bmatrix}^T$ and the input vector has dimensions of $\mathbf{u} \sim \begin{bmatrix} \text{m}/\text{s}^2 & \text{m}/\text{s}^2 \end{bmatrix}^T$. The cost function is defined by $Q = Q_N = \text{diag}\left( \begin{bmatrix} 1 & 1 & 0 & 0 \end{bmatrix} \right)$ and $R = 10 I$. The initial condition is $\mathbf{x}_0 = \begin{bmatrix} 0 & -10 & 0 & 0 \end{bmatrix}^T$ and the discrete time step is $\Delta t = 1.0$~s. The velocity constraint set $\mathcal{V}$ is defined according to $v_{\text{max}} = 5$~m/s and the input constraint set $\mathcal{U}$ is defined according to $v_{\text{min}}=0.1$~m/s and $\omega_{\text{max}}=75$~deg/s. The MPC horizon is $N=55$. In this example, $\mathcal{U}$ and $\mathcal{S}_k$ are centrally symmetric and as such are represented as zonotopes.

Algorithm~\ref{alg:mpc-reachability} is used to build the MPC optimization problem, and Algorithm~\ref{alg:admm} is used to solve it. The optimal trajectory for this example is depicted as red dots in Fig.~\ref{fig:mpc-traj}. Note that this trajectory corresponds to a single solution of the MPC optimization problem (i.e., the MPC is not implemented in closed loop in this example).

\begin{figure}[t]
    \centering
    \input{figs/solution_method_comparison.pgf}
    \caption{MPC solution times versus MPC horizon using an Ubuntu 22.04 desktop with an Intel\textregistered~Core\textsuperscript{\tiny TM} i7-14700 × 28 processor and 32~GB of RAM. CZ denotes a constrained zonotope problem formulation (Algorithm~\ref{alg:mpc-reachability}) while H-rep denotes a sparse MPC problem formulation using H-rep polytopic constraints (Eq.~\eqref{eq:hrep-mpc}). All solvers are invoked via their respective Python interfaces.}
    \label{fig:solve-times}
\end{figure}

\begin{figure}[t]
    \centering
    \input{figs/closed_loop_solution_time_comparison.pgf}
    \caption{\revmargin{4}{4.3/4.4} \protect\revres{4}{MPC solution time statistics for closed-loop simulation. In this box and whisker plot, the medians are depicted with black lines, the boxes enclose the middle 50\% of data points, and the whiskers enclose all data points.}}
    \label{fig:closed-loop-solve-times}
\end{figure}

To benchmark the computational efficiency of our approach, we consider solution time comparisons to state-of-the-art QP solvers OSQP \cite{stellato2020osqp} and Gurobi \cite{gurobi}. OSQP implements a different ADMM variant from that in this paper, while Gurobi implements an interior point solution methodology for QPs by default. OSQP was configured to use default settings with the exception of solution tolerances, which were set to $\epsilon_{\text{abs}}=0.01$ and $\epsilon_{\text{rel}}=0$. These settings were selected to facilitate more direct comparison with our ADMM implementation. Correspondingly, the convergence tolerance for the ZonoOpt ADMM solver was set to $\epsilon_p=\epsilon_d=0.01$ and the solver was configured to use an infinity norm convergence criterion, mirroring OSQP's implementation. Gurobi was configured to use default settings.   

Additionally, we consider comparisons to the following equivalent and widely used MPC optimization problem formulation using H-rep polytopic constraints: 
\begin{subequations} \label{eq:hrep-mpc}
\begin{align}
    &\begin{bmatrix}
        -I & 0 & 0 & \dots & 0 & 0 & 0 \\
        A & B & -I & \cdots & 0 & 0 & 0 \\
        &&&\ddots&&&\\
        0 & 0 & 0 & \dots & A & B & -I
    \end{bmatrix} \begin{bmatrix} \mathbf{x}_0 \\ \mathbf{u}_0 \\ \mathbf{x}_1 \\ \vdots \\ \mathbf{x}_{N-1} \\ \mathbf{u}_{N-1} \\ \mathbf{x}_N \end{bmatrix} = \begin{bmatrix}
        -\mathbf{x}_0 \\ \mathbf{0} \\ \vdots \\ \mathbf{0}
    \end{bmatrix} \;, \\
    &\begin{bmatrix}
        0 & A_u & 0 & \dots & 0 & 0 & 0 \\
        0 & 0 & A_{x,1} & \dots & 0 & 0 & 0 \\
        &&&\ddots&&&\\
        0 & 0 & 0 & \dots & A_{x, N-1} & 0 & 0 \\
        0 & 0 & 0 & \dots & 0 & A_u & 0 \\
        0 & 0 & 0 & \dots & 0 & 0 & A_{x,N} \\
    \end{bmatrix} \begin{bmatrix} \mathbf{x}_0 \\ \mathbf{u}_0 \\ \mathbf{x}_1 \\ \vdots \\ \mathbf{x}_{N-1} \\ \mathbf{u}_{N-1} \\ \mathbf{x}_N \end{bmatrix} \nonumber \\
    &\;\;\; \leq \begin{bmatrix}
        \mathbf{b}_u^T & \mathbf{b}_{x,1}^T & \dots & \mathbf{b}_{x,N-1}^T & \mathbf{b}_u^T & \mathbf{b}_{x,N}^T
    \end{bmatrix}^T \;, \\
    &P = \text{blkdiag}\left( \begin{bmatrix}
        Q & R & Q & \dots & Q & R & Q_N
    \end{bmatrix} \right) \;, \\
    &\mathbf{q} = \left[ \begin{matrix}
        \mathbf{0}^T & \mathbf{0}^T & -(Q \mathbf{x}^r_1)^T & \dots \end{matrix} \right. \nonumber \\
        &\quad \left. \begin{matrix}& -(Q \mathbf{x}^r_{N-1})^T & \mathbf{0}^T & -(Q \mathbf{x}^r_N)^T
    \end{matrix} \right] ^T \;.
\end{align}
\end{subequations}
Eq.~\eqref{eq:hrep-mpc} exploits sparsity for efficient solution and improved scalability with respect to MPC horizon $N$ when compared to dense H-rep MPC optimization problem formulations \cite{borrelli2017predictive, jerez2011condensed}.

\revmargin{3}{3.2}\revres{3}{When Algorithm~\ref{alg:mpc-reachability} is used to build the MPC problem, the $M$ matrix~\eqref{eq:M-def} has $12315$ non-zero elements and there are $825$ optimization variables $\bm{\xi}$. Using \eqref{eq:hrep-mpc} to build the MPC problem, an equivalent ADMM formulation would require an $M$ matrix with $13052$ elements and $1984$ optimization variables. See Sec.~\ref{sec:admm-formulation} for further discussion of ADMM complexity using a constrained zonotope constraint representation.}

Solution times for the MPC optimization problem are compared for the considered QP solvers (ZonoOpt, OSQP, and Gurobi) and problem formulations (Algorithm~\ref{alg:mpc-reachability} vs. \eqref{eq:hrep-mpc}) in Fig.~\ref{fig:solve-times}. To assess the scalability of these solution approaches to large MPC horizons, the MPC problem was modified using a scaling factor $f$ such that $N = fN_0$ and $\Delta t = \Delta t_0/f$ where $N_0 = 55$ and $\Delta t_0 = 1$~s. The position constraint sets $\mathcal{P}_k$ and reference states $\mathbf{x}^r_k$ are correspondingly adjusted via interpolation. The MPC optimization problem was solved for $f = \{1, 2, ..., 21 \}$, resulting in MPC horizons ranging from $N=55$ to $N=1155$. 

\revmargin{4}{4.3/4.4}\revres{4}{We additionally simulate the MPC in closed loop for the problem depicted in Fig.~\ref{fig:mpc-traj}. Here, the scaling factor was set to $f=4$ and the system was simulated for $220$ time steps. The receding horizon length was $N=100$. MPC solution time statistics over the course of the simulation for each solution approach are shown in Fig.~\ref{fig:closed-loop-solve-times}.}

Figs.~\ref{fig:solve-times} and \ref{fig:closed-loop-solve-times} empirically show that the proposed MPC problem formulation using reachability analysis of constrained zonotopes and corresponding solution approach using ADMM achieves lower solution times \revmargin{4}{4.3/4.4}\revres{4}{on average} and improved scalability with respect to MPC horizon when compared to state-of-the-art problem formulations and QP solvers. \revmargin{4}{4.4} \revres{4}{These efficiency improvements are a result of the efficient use of the constrained zonotope structure in the ADMM iterations.}

\subsection{Moving Horizon Estimation and Set-Valued State Estimation} \label{sec:results-mhe}

Consider the combined MHE and SVSE problem described in Sec.~\ref{sec:app-mhe} with a noise-perturbed double integrator model where $A$, $B$, $\mathbf{x}_k$, and $\mathbf{u}_k$ are defined as in \eqref{eq:results-dbl-int-dyn} and have the same dimensions as in Sec.~\ref{sec:results-mpc}. The measurement matrix is $C=I$. The discrete time step is $\Delta t=1$~s. The set of possible states is given as $\mathcal{S} = \mathcal{S}_p \times \mathcal{S}_v$ where $\mathcal{S}_p$ and $\mathcal{S}_v$ are both regular hexagons. The inner radii of $\mathcal{S}_p$ and $\mathcal{S}_v$ are 500~m and 1~m/s, respectively. The process and measurement noise similarly have the forms $\mathcal{W} = \mathcal{W}_p \times \mathcal{W}_v$ and $\mathcal{V} = \mathcal{V}_p \times \mathcal{V}_v$ where $\mathcal{W}_p$, $\mathcal{V}_p$, $\mathcal{W}_v$, and $\mathcal{V}_v$ are regular hexagon approximations of circular bounds on position and velocity noise distributions.
Truncated normal distributions are used such that the position and velocity noise sets have inner radii of $2 \sigma$. Zero-mean noise is used in all cases. The position process noise has standard deviation $\sigma=0.001$~m, and the velocity process noise has standard deviation $\sigma=0.01$~m/s. The position measurement noise has standard deviation $\sigma=0.5$~m, and the velocity measurement noise has standard deviation $\sigma=0.2$~m/s. The process and measurement noise covariance matrices $Q$ and $R$ are set accordingly. The initial state covariance matrix is set as $P_{-N}^{-1}=0$, which is a common selection \cite{rao2003constrained, muske1995nonlinear}. The set of possible states at simulation time step $k=0$ is $\mathcal{X}_{k=0} = \mathcal{X}_{p0} \times \mathcal{X}_{v0}$ where $\mathcal{X}_{p0}$ and $\mathcal{X}_{v0}$ are again regular hexagons. The initial position state set has inner radius 2~m and is centered at $(x,y) = (-4,1)$. The initial velocity state set has inner radius 1~m/s and is centered at $(\dot{x}, \dot{y})=(0,0)$. The true initial state is $\mathbf{x}_0 = \begin{bmatrix} -5 & 2 & -0.6 & 0.2 \end{bmatrix}^T$. The moving horizon length is $N=15$. For simulation time steps less than $k=N$, the horizon length is set to $k$, and $\mathcal{X}_{-N}=\mathcal{X}_{k=0}$. A pre-determined sequence of inputs $\mathbf{u}_k$ is applied to the system. Simulated noise inputs are sampled from the previously described truncated normal distributions. The simulation length is 40 time-steps. 

Simulated states, MHE estimates, and feasible sets for the numerical example are given in Fig.~\ref{fig:mhe-traj}, with \revres{5}{Algorithm~\ref{alg:mhe}}\revmargin{5}{5.5} used to compute the MHE estimates and feasible sets. The left sub-figure depicts the position results while the right sub-figure depicts the velocity results. The RMS position error from the measurements is 0.505~m and the RMS velocity error is 0.200~m/s. The RMS position error from the MHE solution is 0.224~m and the RMS velocity error is 0.066~m/s. Using the computer described in Fig.~\ref{fig:solve-times},  
\revres{4}{the mean solution time for this example was $0.27$~ms (min: $0.05$~ms, max: $0.58$~ms).}\revmargin{4}{4.5}
The true state and MHE estimate always lie within the feasible set while the measurement may lie outside the feasible set. In this example, the widths of the feasible sets are generally much larger than the MHE RMS errors.
By combining SVSE and MHE as has been done here, one can compute an optimal state estimate in the sense of \eqref{eq:mhe} while simultaneously computing the feasible set $\mathcal{X}_0$ that contains all possible states of the system.

\begin{figure*}
    \centering
    \input{figs/mhe_traj.pgf}
    \caption{Simulated position and velocity estimates and feasible sets \revmargin{5}{5.6}\protect\revres{5}{for a double integrator system} using combined SVSE and MHE. \protect\revres{5}{Given an initial set containing the state of the system, the blue feasible sets are guaranteed to contain the system state. The MHE provides a best estimate of the system state within the blue feasible sets.}}
    \label{fig:mhe-traj}
\end{figure*}

\subsection{Safety Verification} \label{sec:results-reachability}

Consider the safety verification problem described in Sec.~\ref{sec:app-safety} using a disturbed discrete-time double integrator system with $A$ and $B$ defined by \eqref{eq:results-dbl-int-dyn}. 
\revmargin{5}{5.4}\revres{5}{This scenario is depicted in Fig.~\ref{fig:reach-traj}}.
The unsafe set $\mathcal{O}$ only contains position states, so safety verification at time step $k$ entails verifying the infeasibility of $\mathcal{Z} = \mathcal{X}_k \cap_{\begin{bmatrix} I & 0 \end{bmatrix}} \mathcal{O}$. In ZonoOpt, this feasibility check is implemented by executing Algorithm~\ref{alg:admm} with $\tilde{P}=I$ and $\mathbf{q} = \mathbf{0}$. 

In this example, the discrete time step is $\Delta t =0.5$~s. The state feedback gain $K$ is computed as the discrete LQR gain with quadratic cost function matrices $Q = \text{diag}\left(\begin{bmatrix}1 & 1 & 0 & 0 \end{bmatrix}\right)$ and $R = 0.1 I$. The set of feasible states $\mathcal{S}$ is the same as that presented in Sec.~\ref{sec:results-mhe}. The disturbance set $\mathcal{W}$ is defined as $\mathcal{W} = \mathcal{W}_p \times \mathcal{W}_v$ with $\mathcal{W}_p$ and $\mathcal{W}_v$ regular hexagons of radius 0.01~m and 0.2~m/s, respectively. The center of $\mathcal{W}_v$ is $(\dot{x}, \dot{y}) = (0, 0.5)$ while $\mathcal{W}_p$ is centered at $\mathbf{0}$. The initial state set $\mathcal{X}_0$ is given as $\mathcal{X}_0 = \mathcal{X}_{p0} \times \mathcal{X}_{v0}$ where $\mathcal{X}_{p0}$ and $\mathcal{X}_{v0}$ are regular hexagons of radius 0.5~m and 0.5~m/s, respectively, and with centers $(x,y) = (1,0)$ and $(\dot{x}, \dot{y}) = (0,0)$. The infeasibility checking interval in Algorithm~\ref{alg:admm} is set to $k_{\text{inf}}=1$.

Fig.~\ref{fig:reach-traj} shows the results for this numerical example.
Algorithm~\ref{alg:admm} finds a certificate of infeasibility for $\mathcal{X}_k \cap_{\begin{bmatrix} I & 0 \end{bmatrix}} \mathcal{O}$ for all time steps $k$, indicating that trajectories of the system are guaranteed to be safe. The sets $\mathcal{X}_k$ are color coded to show how many ADMM iterations were required to generate the infeasibility certificate. In the majority of cases, only a single iteration is required. Using the desktop computer described in Fig.~\ref{fig:solve-times},  
\revres{4}{the mean time to produce an infeasibility certificate was $0.058$~ms (min: $0.024$~ms, max: $0.129$~ms).}\revmargin[true]{4}{4.6}

\begin{figure}[t]
    \centering
    \input{figs/reachability_traj.pgf}
    \caption{Safety verification for the disturbed double integrator system with LQR state feedback control. \revmargin{5}{5.6/5.7}\protect\revres{5}{The $\mathcal{X}_k$ are the projections of the reachable sets of the system, $\mathcal{O}$ is the unsafe set, and the $\mathbf{x}^r_k$ are the state references. Thm.~\ref{thm:inf-certificate} is used to prove that there exists no trajectory of the system that intersects the unsafe set in a handful of ADMM iterations.}}
    \label{fig:reach-traj}
\end{figure}
\section{Experimental Results} \label{sec:experiment}

\revres{6}{\subsection{Experimental Setup}}\revmargin{6}{6.9}
A robotics experiment was conducted to demonstrate the effectiveness of the proposed methods in a real-time context using resource-constrained, embedded hardware. This experiment implemented Algorithm~\ref{alg:mhe} for combined SVSE and MHE for robot localization. A Husarion ROSbot 2R was used as the robot platform. All SVSE/MHE calculations were executed onboard the robot's Raspberry Pi 4, \revres{4}{which has a quad-core ARM-8 Cortex-A72 @ 1.5 GHz processor and 4GB RAM.}\revmargin[true]{4}{4.7} Optitrack Prime$^\text{X}$22 motion capture cameras were used as the ground truth source for the robot's position. Teleoperation was used to drive the robot around the laboratory space. Data from the  robot's inertial measurement unit (IMU), wheel motor encoders, and corrupted measurements from the motion capture system were fused to generate a best estimate of the robot's motion states and accompanying bounding set. All algorithms were executed within a ROS2~\revres{6}{Humble}\revmargin[true]{6}{6.10} framework, and the SVSE/MHE was implemented as a C++ ROS2 node using the ZonoOpt toolbox.

\revres{6}{\subsection{Implementation Details}}\revmargin{6}{6.9}
The ROSbot 2R is a differential drive robot described by the equations of motion
\begin{subequations}
\begin{align}
    &v = \frac{d}{4} (\omega_R + \omega_L) \;,\; \omega = \frac{d}{2l} (\omega_R - \omega_L) \;, \label{eq:robot-dyn-diff-drive} \\
    &\dot{x} = v \cos{(\theta)} \;,\; \dot{y} = v \sin{(\theta)} \;,\; \dot{\theta} = \omega \;, \label{eq:robot-dyn-unicycle}
\end{align}
\end{subequations}
where $v$ is linear velocity, $\omega$ is angular velocity, $\theta$ is the heading angle, and $x$ and $y$ are position coordinates. The angular velocities of the right and left wheels are $\omega_R$ and $\omega_L$, respectively, $d$ is the wheel diameter, and $l$ is the distance between the right and left wheels. Note that~\eqref{eq:robot-dyn-unicycle} is the unicycle model, which is differentially flat in $x$ and $y$ and can be described using the double integrator model~\cite{diffflat}. As such, the SVSE/MHE uses the discrete-time double integrator model~\eqref{eq:results-dbl-int-dyn}.

The control inputs for this model are the lab frame acceleration components $\mathbf{u} = \begin{bmatrix} \tilde{a}_x & \tilde{a}_y \end{bmatrix}^T$. For the experiment, we define the measurements to be the lab frame position and velocity components $\mathbf{y} = \begin{bmatrix}
    \tilde{x} & \tilde{y} & \tilde{v}_x & \tilde{v}_y
\end{bmatrix}^T$. The measurement matrix is $C=I$. The tilde notation is used here to indicate a measured quantity. The inputs and measurements are derived from motion capture position feedback, wheel motor encoder feedback, and IMU feedback.

The IMU directly senses body frame acceleration $\tilde{\mathbf{a}}^b = \begin{bmatrix} \tilde{a}^b_{x} & \tilde{a}^b_{y} \end{bmatrix}^T$ and angular velocity. The IMU also outputs an orientation quaternion based on the integrated angular velocity. The robot's heading angle in the lab frame, $\tilde{\theta}$, is derived from this orientation quaternion.

The wheel motor encoders provide wheel angular velocity feedback. The two right wheel angular velocities are averaged to give $\tilde{\omega}_R$, and the two left wheel angular velocities are averaged to give $\tilde{\omega}_L$. The body frame velocity measurement is then computed as $\tilde{\mathbf{v}}^b = \begin{bmatrix} \tilde{v} & 0 \end{bmatrix}^T$ where $\tilde{v}$ is computed from $\tilde{\omega}_R$ and $\tilde{\omega}_L$ using~\eqref{eq:robot-dyn-diff-drive}.

The motion capture system provides high-accuracy position feedback signals $\mathbf{x}_p = \begin{bmatrix}
    x & y
\end{bmatrix}^T$. To simulate less precise position feedback such as that which might be provided by GPS, random noise drawn from a $\pm 2\sigma$-bounded Gaussian distribution was added to the motion capture feedback. These corrupted position feedback signals are denoted as $\tilde{\mathbf{x}}_p = \begin{bmatrix} \tilde{x} & \tilde{y} \end{bmatrix}^T$, and the position measurement uncertainty standard deviation is given as $\sigma_p$. 

Transforming body frame quantities into the lab frame, the input and measurement vectors used in the SVSE/MHE are
\begin{subequations}
\begin{align}
    &\mathbf{u} = \begin{bmatrix}
        \tilde{a}_x \\ \tilde{a}_y
    \end{bmatrix} = \begin{bmatrix}
        \cos(\tilde{\theta}) & -\sin(\tilde{\theta}) \\
        \sin(\tilde{\theta}) & \cos(\tilde{\theta})
    \end{bmatrix} \begin{bmatrix}
        \tilde{a}^b_x \\ \tilde{a}^b_y
    \end{bmatrix} \;, \\
    &\mathbf{y} = \begin{bmatrix}
        \tilde{x} \\ \tilde{y} \\ \tilde{v}_x \\ \tilde{v}_y
    \end{bmatrix} = \begin{bmatrix}
        1 & 0 & 0 & 0 \\
        0 & 1 & 0 & 0 \\
        0 & 0 & \cos(\tilde{\theta}) & -\sin(\tilde{\theta}) \\
        0 & 0 & \sin(\tilde{\theta}) & \cos(\tilde{\theta})
    \end{bmatrix} \begin{bmatrix}
        \tilde{x} \\ \tilde{y} \\ \tilde{v} \\ 0
    \end{bmatrix} \;.
\end{align}
\end{subequations}

Due to uncertainty in the sensed heading angle $\tilde{\theta}$, disturbance and measurement uncertainty sets $\mathcal{W}$ and $\mathcal{V}$ must account for errors incurred when translating sensed acceleration and velocity from the body frame into the lab frame. 
In this implementation, these errors were bounded using interval arithmetic. \revres{4}{The following calculations are analagous to methods presented in~\cite{alamo2005guaranteed}.}\revmargin{4}{4.9}
The true heading angle $\theta$ was assumed to belong to the interval set $\theta \in [\theta] \equiv [\tilde{\theta} - e_{\theta}, \tilde{\theta} + e_{\theta}]$, where $e_{\theta}$ is an error bound. Bounds on $\cos (\theta)$ and $\sin (\theta)$ over the interval $[\theta]$ are then defined as $[c \theta]$ and $[s \theta]$. Error intervals for the trigonometric functions are then $[e_{c \theta}] = [c \theta] - [\cos (\tilde{\theta}), \cos(\tilde{\theta})]$ and $[e_{s \theta}] = [s \theta] - [\sin (\tilde{\theta}), \sin (\tilde{\theta})]$.

The acceleration uncertainty set in the lab frame is given as 
\begin{equation} \label{eq:W_u-calc}
\mathcal{W}_u = \begin{bmatrix} \cos (\tilde{\theta}) & -\sin (\tilde{\theta}) \\ \sin (\tilde{\theta}) & \hphantom{-}\cos (\tilde{\theta}) \end{bmatrix} \mathcal{W}_u^b \oplus \mathcal{E}_u \;,
\end{equation}
where $\mathcal{W}_u^b$ is the acceleration measurement error set in the body frame and $\mathcal{E}_u$ accounts for heading angle uncertainty. In this case, $\mathcal{W}_u^b \equiv [\mathbf{e}_u^b]$ where $ [\mathbf{e}_u^b] = \begin{bmatrix} [e_{a,x}^b] & [e_{a,y}^b] \end{bmatrix}^T$
is a box, and $[e^b_{a,x}]$ and $[e^b_{a,y}]$ are the body frame acceleration error intervals. An over-approximation of the uncertainty in $\mathcal{W}_u$ due to heading angle uncertainty is
\begin{equation}
    \mathcal{E}_u = \begin{bmatrix}
        [e_{c\theta}][e_{a,x}^b] - [e_{s \theta}][e_{a,y}^b] \\
        [e_{s\theta}][e_{a,x}^b] + [e_{c \theta}][e_{a,y}^b]
    \end{bmatrix} \;.
\end{equation}
To implement~\eqref{eq:W_u-calc}, $\mathcal{W}_u^b$ and $\mathcal{E}_u$ were first converted to zonotopes, and then the identities in~\eqref{eq:set-ops} were used to compute $\mathcal{W}_u$. The disturbance set $\mathcal{W}$ is then given as $\mathcal{W} = B \mathcal{W}_u$. 

The measurement error set is given as $\mathcal{V} = \mathcal{V}_p \times \mathcal{V}_v$, where $\mathcal{V}_p$ is the position measurement error set and $\mathcal{V}_v$ is the velocity measurement error set. The position measurement error set is 
\begin{equation}
    \mathcal{V}_p = \begin{bmatrix}
        \sigma_p & 0 \\
        0 & \sigma_p
    \end{bmatrix} \mathcal{Z}_{\circ} \;,
\end{equation}
where $\mathcal{Z}_{\circ}$ is an outer approximation of a unit circle, taken here as an eight-sided zonotope, \revres{4}{which over-approximates the area of the set by about $5\%$.}\revmargin{4}{4.10}
The velocity measurement error set $\mathcal{V}_v$ was computed analogously to $\mathcal{W}_u$, with the body frame velocity error set given as $\mathcal{V}^b_v \equiv [\mathbf{e}_v^b]$ where $[\mathbf{e}_v^b] = \begin{bmatrix}
    [-e_v, e_v] & [-\epsilon, \epsilon]
\end{bmatrix}^T$. In this case, $e_v$ is the maximum error in the velocity derived from the wheel motor encoder measurements, and $0 < \epsilon \ll 1$. 

For the MHE, covariance matrices $Q$ and $R$ were estimated by transforming body frame covariance matrices into the lab frame as required for velocity and acceleration. Standard deviations are taken to be one half of the uncertainty bound, e.g., the body frame forward velocity standard deviation is $\sigma_v = \frac{1}{2} e_v$. The covariance matrices do not account for heading angle uncertainty contributions. \remove{As in Sec.~\ref{sec:results-mhe}, }\revmargin[true]{5}{5.3}{The initial inverse covariance matrix was taken to be $P^{-1}_N = 0$.

The set of possible states is $\mathcal{S} = \mathcal{S}_p \times \mathcal{S}_v$, where $\mathcal{S}_p$ and $\mathcal{S}_v$ are the possible position and velocity states. The possible position states are defined by the dimensions of the laboratory space, and the possible velocity states are given as 
\begin{equation}
    \mathcal{S}_v = \begin{bmatrix}
        v_{\text{max}} & 0 \\
        0 & v_{\text{max}}
    \end{bmatrix} \mathcal{Z}_{\circ} \;,
\end{equation}
where $v_{\text{max}}$ is a bound on the translational velocity.

Position feedback was published at a rate of $100$~Hz, and the robot published both the IMU and wheel motor encoder feedback messages at $20$~Hz. The SVSE/MHE ran at a rate of $10$~Hz and was triggered off of the IMU message, i.e., the SVSE/MHE algorithm executed immediately following receipt of the IMU message, every other IMU message. The horizon length was $N=20$. The initial feasible set was computed from the initial measurements as 
\begin{equation}
    \mathcal{X}_0 = \begin{bmatrix}
        \tilde{x}_0 & \tilde{y}_0 & \tilde{v}_{x,0} & \tilde{v}_{y,0}
    \end{bmatrix}^T \oplus \mathcal{V} \;.
\end{equation}

To mitigate complexity growth of the initial feasible set $\mathcal{X}_{-N}$ over repeated SVSE/MHE iterations, $\mathcal{X}_{-N}$ was periodically replaced with its bounding box approximation. The bounding box was computed by calculating the support, $\text{sup}(\mathcal{X}_{-N}, \mathbf{d}) = \max_{\mathbf{z} \in \mathcal{X}_{-N}} \langle \mathbf{z}, \mathbf{d} \rangle$, along each axis-aligned direction $\mathbf{d}$. The resulting box is then converted to a zonotope. Here, each bounding box calculation requires eight invocations of Algorithm~\ref{alg:admm}. The factorizations of $M$ and $A A^T$ were reused between optimizations for efficiency. Bounding box approximations were computed every 10 SVSE/MHE iterations following execution of Algorithm~\ref{alg:mhe}.

\begin{table}
    \caption{Parameters for SVSE/MHE experimental implementation.}
    \centering
    \begin{tabular}{c|c c c c c c c}
        \toprule
         Param. & $d$ & $v_{\text{max}}$ & $\sigma_p$ & $e_{\theta}$ & $e_v$ & $e^b_{u,x}$ & $e^b_{u,y}$ \\ \midrule 
         Units & m & m/s & m & rad & m/s & m/s$^2$ & m/s$^2$ \\ \midrule
         Value & 0.084 & 1.1 & 0.3 & 0.262 & 0.25 & 3.0 & 3.0
    \end{tabular}
    \label{tab:svse-mhe-params}
\end{table}

Parameters for the experimental implementation are given in Table~\ref{tab:svse-mhe-params}. The heading uncertainty $e_{\theta}$ accounts for both initialization error and errors in the IMU's heading computation. The velocity uncertainty $e_v$ is dominated by errors due to asynchronous sampling between the IMU and wheel motor encoder measurements. Acceleration uncertainty bounds were estimated from inspection of recorded acceleration data.

\begin{figure*}[t]
    \centering
    \input{figs/svse_mhe_experiment.pgf}
    \caption{Experimental implementation of combined MHE and SVSE for robot localization. Feasible sets $\mathcal{X}_k$, motion capture positions $\mathbf{x}_{p,k}$, measurements $\mathbf{y}_{p,k}$, and MHE estimates $\hat{\mathbf{x}}_{p,k}$ are overlaid on top of corrected images from an overhead fisheye camera. Due to imperfect correction of the fisheye effect and differences in sampling rate, images of the robot do not perfectly coincide with the motion capture positions $\mathbf{x}_{p,k}$.}
    \label{fig:svse-mhe-experiment}
\end{figure*}

\begin{figure}
    \centering
    \input{figs/svse_mhe_experiment_timeseries.pgf}
    \caption{\revmargin{4}{4.12}\protect\revres{4}{Time series plot of motion capture positions $\mathbf{x}_{p,k}$, position measurements $\mathbf{y}_{p,k}$, MHE estimates $\hat{\mathbf{x}}_{p,k}$, and SVSE bounds $\mathcal{X}_k$ for the first 75 time steps of the combined MHE and SVSE experiment.}}
    \label{fig:svse-mhe-exp-timeseries}
\end{figure}

\revres{6}{\subsection{Results}}\revmargin{6}{6.9}
Results for this experiment are displayed in Fig.~\ref{fig:svse-mhe-experiment}. Here, the MHE position estimates $\hat{\mathbf{x}}_{p,k}$ and projections of the SVSE bounding sets $\mathcal{X}_k$ onto the position states are overlaid on top of overhead camera images of the laboratory space. The motion capture positions are given by $\mathbf{x}_{p,k}$ and are taken to be ground truth. The position measurement feedback for the SVSE/MHE is given by $\mathbf{y}_{p,k}$. Because there is significant overlap between adjacent $\mathcal{X}_k$, only every tenth iteration is plotted in Fig.~\ref{fig:svse-mhe-experiment}.
\revmargin[false]{4}{4.12}\revres{4}{A time-series plot of the same data is given in Fig.~\ref{fig:svse-mhe-exp-timeseries}. For readability, only the first 75 time steps (of 400) are displayed.}
Effectiveness of the SVSE is verified by the observation that $\mathbf{x}_{p,k} \in [I \; 0]\mathcal{X}_k \; \forall k$. The MHE significantly improved position estimation accuracy when compared to the position measurements. \revres{4}{The RMS error for the position measurements $\mathbf{y}_{p,k}$ was 0.31~m, while the RMS error for the MHE position estimates $\hat{\mathbf{x}}_{p,k}$ was 0.11~m.}\revmargin{4}{4.12}

\revmargin{4}{4.13}\revres{4}{Problem formulation and solution time statistics for the SVSE/MHE experiment are given in Tab.~\ref{tab:svse-mhe-results}. The MHE converged quickly in this experiment. This is because the optimal estimate was generally in the interior of the SVSE feasible set, so $\bm{\xi}_k$ iterates~\eqref{eq:admm-implementation-xi} were typically feasible and few ADMM iterations were needed to achieve dual convergence.
The main computational burden in this experiment was computing the bounding box over-approximations, as these each required eight invocations of Alg.~\ref{alg:admm}, and optimal solutions were on the boundary of the SVSE feasible set.}

\setlength{\tabcolsep}{7pt}
\begin{table}
    \caption{\revmargin{4}{4.13}\protect\revres{4}{Problem formulation and solution time statistics for SVSE/MHE experiment.}}
    \centering
    \begin{tabular}{l|c|c c c}
        \toprule
         & unit & median & min & max \\ \midrule
         SVSE (Alg.~\ref{alg:mhe}) time & ms & 8.58 & 0.067 & 13.35 \\ \midrule
         MHE (Alg.~\ref{alg:admm}) time & ms & 2.06 & 0.20 & 3.71 \\ \midrule
         MHE (Alg.~\ref{alg:admm}) iterations & counts & 2 & 1 & 4 \\ \midrule
         Bounding box time & ms & 49.99 & 31.54 & 69.69 \\
         \end{tabular}
    \label{tab:svse-mhe-results}
\end{table}
\section{Conclusion} \label{sec:conclusion}
This paper considers the problem of solving QPs for which the constraints are represented as constrained zonotopes. Novel reachability computations were proposed that produce constrained zonotopes for which the generator and constraint matrices have much higher sparsity than existing methods. An ADMM algorithm was proposed to efficiently solve the QPs. 
The numerical results demonstrate that when using the sparse reachability calculations to build an MPC problem, the ADMM algorithm solves the resulting QP faster than state-of-the-art QP solvers and a sparse MPC problem formulation, with the solution time gap increasing as the MPC horizon increases. Additionally, reachability computations were used to build a state estimator that combines MHE with SVSE, and this state estimator was experimentally applied for online robot localization. Efficient safety verification of system trajectories using ADMM infeasibility detection was demonstrated as well.  

Independent of the ADMM algorithm described in this paper, the sparse reachability calculations can be used to reduce the memory complexity of constrained zonotope-based reachability computations when compared to existing methods. Furthermore, analysis of the reachable sets is likely to be accelerated when using any optimization algorithm based on sparse linear algebra.

This paper highlights that reachability calculations with zonotopes and constrained zonotopes can be used to build optimal control and estimation problems amenable to efficient optimization.
To facilitate workflows based on real-time reachability analysis and optimization, the open-source toolbox ZonoOpt was developed.

\bibliography{bibitems}

@book{borrelli2017predictive,
  title={Predictive Control for Linear and Hybrid Systems},
  author={Borrelli, Francesco and Bemporad, Alberto and Morari, Manfred},
  year={2017},
  publisher={Cambridge University Press}
}

@MISC{eigenweb,
  author = {Ga\"{e}l Guennebaud and Beno\^{i}t Jacob and others},
  title = {Eigen v3},
  howpublished = {http://eigen.tuxfamily.org},
  year = {2010}
 }

@inproceedings{koeln2023zonolab,
  title={{zonoLAB}: A {MATLAB} Toolbox for Set-Based Control Systems Analysis Using Hybrid Zonotopes},
  author={Koeln, Justin and Bird, Trevor J and Siefert, Jacob and Ruths, Justin and Pangborn, Herschel and Jain, Neera},
  booktitle={American Control Conference},
  pages={2498-2505},
  year={2024}
}

@article{scott2016constrained,
  title={Constrained zonotopes: A new tool for set-based estimation and fault detection},
  author={Scott, Joseph K and Raimondo, Davide M and Marseglia, Giuseppe Roberto and Braatz, Richard D},
  journal={Automatica},
  volume={69},
  pages={126--136},
  year={2016},
  publisher={Elsevier}
}

@misc{gurobi,
  author = {{Gurobi Optimization, LLC}},
  title = {{Gurobi Optimizer Reference Manual}},
  year = 2023,
  url = "https://www.gurobi.com"
}

@book{ziegler2012lectures,
  title={Lectures on Polytopes},
  author={Ziegler, G{\"u}nter M},
  volume={152},
  year={2012},
  publisher={Springer Science \& Business Media}
}

@article{boyd2011distributed,
  title={Distributed optimization and statistical learning via the alternating direction method of multipliers},
  author={Boyd, Stephen and Parikh, Neal and Chu, Eric and Peleato, Borja and Eckstein, Jonathan and others},
  journal={Foundations and Trends in Machine learning},
  volume={3},
  number={1},
  pages={1--122},
  year={2011},
  publisher={Now Publishers, Inc.}
}

@article{raghunathan2014admm,
  title={{ADMM} for convex quadratic programs: Q-linear convergence and infeasibility detection},
  author={Raghunathan, Arvind U and Di Cairano, Stefano},
  journal={arXiv preprint arXiv:1411.7288},
  year={2014}
}

@inproceedings{rey2016admm,
  title={{ADMM} prescaling for model predictive control},
  author={Rey, Felix and Frick, Damian and Domahidi, Alexander and Jerez, Juan and Morari, Manfred and Lygeros, John},
  booktitle={IEEE 55th Conference on Decision and Control},
  pages={3662--3667},
  year={2016}
}

@article{rey2020admm,
  title={{ADMM} for exploiting structure in {MPC} problems},
  author={Rey, Felix and Hokayem, Peter and Lygeros, John},
  journal={IEEE Transactions on Automatic Control},
  volume={66},
  number={5},
  pages={2076--2086},
  year={2020},
}

@book{nocedal1999numerical,
  title={Numerical Optimization},
  author={Nocedal, Jorge and Wright, Stephen J},
  year={1999},
  publisher={Springer}
}

@book{boyd2004convex,
  title={Convex Optimization},
  author={Boyd, Stephen P and Vandenberghe, Lieven},
  year={2004},
  publisher={Cambridge University Press}
}

@inproceedings{chari2024constraint,
  title={Constraint preconditioning and parameter selection for a first-order primal-dual method applied to model predictive control},
  author={Chari, Govind M and Yu, Yue and A{\c{c}}{\i}me{\c{s}}e, Beh{\c{c}}et},
  booktitle={IEEE 63rd Conference on Decision and Control},
  pages={1676--1683},
  year={2024}
}

@article{stellato2020osqp,
  title={{OSQP}: An operator splitting solver for quadratic programs},
  author={Stellato, Bartolomeo and Banjac, Goran and Goulart, Paul and Bemporad, Alberto and Boyd, Stephen},
  journal={Mathematical Programming Computation},
  volume={12},
  number={4},
  pages={637--672},
  year={2020},
  publisher={Springer}
}

@article{srikanthan2024closed,
  title={Closed-Loop Analysis of {ADMM}-Based Suboptimal Linear Model Predictive Control},
  author={Srikanthan, Anusha and Karapetyan, Aren and Kumar, Vijay and Matni, Nikolai},
  journal={IEEE Control Systems Letters},
  year={2024},
}

@inproceedings{raghunathan2014optimal,
  title={Optimal step-size selection in alternating direction method of multipliers for convex quadratic programs and model predictive control},
  author={Raghunathan, Arvind U and Di Cairano, Stefano},
  booktitle={Proceedings of Symposium on Mathematical Theory of Networks and Systems},
  pages={807--814},
  year={2014},
  organization={Citeseer}
}

@article{raghuraman2022set,
  title={Set operations and order reductions for constrained zonotopes},
  author={Raghuraman, Vignesh and Koeln, Justin P},
  journal={Automatica},
  volume={139},
  pages={110204},
  year={2022},
  publisher={Elsevier}
}

@article{siefert2025reachability,
  title={Reachability Analysis Using Hybrid Zonotopes and Functional Decomposition},
  author={Siefert, Jacob A and Bird, Trevor J and Thompson, Andrew F and Glunt, Jonah J and Koeln, Justin P and Jain, Neera and Pangborn, Herschel C},
  journal={IEEE Transactions on Automatic Control},
  year={2025},
}

@article{althoff2021set,
  title={Set Propagation Techniques for Reachability Analysis},
  author={Althoff, Matthias and Frehse, Goran and Girard, Antoine},
  journal={Annual Review of Control, Robotics, and Autonomous Systems},
  volume={4},
  number={1},
  pages={369--395},
  year={2021},
  publisher={Annual Reviews}
}

@inproceedings{raghunathan2014infeasibility,
  title={Infeasibility detection in alternating direction method of multipliers for convex quadratic programs},
  author={Raghunathan, Arvind U and Di Cairano, Stefano},
  booktitle={53rd IEEE Conference on Decision and Control},
  pages={5819--5824},
  year={2014}
}

@inproceedings{whitaker2021optimal,
  title={Optimal Path Smoothing while Maintaining a Region of Safe Operation},
  author={Whitaker, Justin and Droge, Greg},
  booktitle={IEEE American Control Conference},
  pages={3830--3835},
  year={2021}
}

@book{diffflat,
    author = {Sira-Ramirez, Hebertt and Agrawal, Sunil K.},
    title = {Differentially Flat Systems},
    publisher = {Marcel Dekker, Inc.},
    year = {2004}
}

@article{zhang2023hierarchical,
  title={A hierarchical multi-vehicle coordinated motion planning method based on interactive spatio-temporal corridors},
  author={Zhang, Xiang and Wang, Boyang and Lu, Yaomin and Liu, Haiou and Gong, Jianwei and Chen, Huiyan},
  journal={IEEE Transactions on Intelligent Vehicles},
  volume={9},
  number={1},
  pages={2675--2687},
  year={2023},
}

@inproceedings{jerez2011condensed,
  title={A condensed and sparse {QP} formulation for predictive control},
  author={Jerez, Juan L and Kerrigan, Eric C and Constantinides, George A},
  booktitle={IEEE 50th Conference on Decision and Control and European Control Conference},
  pages={5217--5222},
  year={2011}
}

@article{rao2003constrained,
  title={Constrained state estimation for nonlinear discrete-time systems: Stability and moving horizon approximations},
  author={Rao, Christopher V and Rawlings, James B and Mayne, David Q},
  journal={IEEE Transactions on Automatic Control},
  volume={48},
  number={2},
  pages={246--258},
  year={2003},
}

@incollection{muske1995nonlinear,
  title={Nonlinear moving horizon state estimation},
  author={Muske, Kenneth R and Rawlings, James B},
  booktitle={Methods of Model Based Process Control},
  pages={349--365},
  year={1995},
  publisher={Springer}
}

@book{jaulin2001interval,
  title={Applied Interval Analysis},
  author={Jaulin, Luc and Kieffer, Michel and Didrit, Olivier and Walter, {\'E}ric},
  year={2001},
  publisher={Springer}
}

@book{rockafellar1997convex,
  title={Convex Analysis},
  author={Rockafellar, R Tyrrell},
  volume={28},
  year={1997},
  publisher={Princeton University Press}
}

@article{robbins2024mixed,
  title={Mixed-Integer {MPC}-Based Motion Planning Using Hybrid Zonotopes with Tight Relaxations},
  author={Robbins, Joshua A and Siefert, Jacob A and Brennan, Sean and Pangborn, Herschel C},
  journal={IEEE Transactions on Control Systems Technology},
  year={2026},
  doi={10.1109/TCST.2026.3661166}
}

@inproceedings{robbins2024efficient,
  title={Efficient solution of mixed-integer {MPC} problems for obstacle avoidance using hybrid zonotopes},
  author={Robbins, Joshua A and Brennan, Sean B and Pangborn, Herschel C},
  booktitle={63rd IEEE Conference on Decision and Control},
  year={2024}
}

@article{srikanthan2023augmented,
  title={Augmented {Lagrangian} methods as layered control architectures},
  author={Srikanthan, Anusha and Kumar, Vijay and Matni, Nikolai},
  journal={arXiv preprint arXiv:2311.06404},
  year={2023}
}

@article{vinod2025projection,
  title={Projection-free computation of robust controllable sets with constrained zonotopes},
  author={Vinod, Abraham P and Weiss, Avishai and Di Cairano, Stefano},
  journal={Automatica},
  volume={175},
  pages={112211},
  year={2025},
  publisher={Elsevier}
}

@article{koeln2020vertical,
  title={Vertical hierarchical {MPC} for constrained linear systems},
  author={Koeln, Justin and Raghuraman, Vignesh and Hencey, Brandon},
  journal={Automatica},
  volume={113},
  pages={108817},
  year={2020},
  publisher={Elsevier}
}

@article{althoff2016cora,
  title={{CORA} 2016 manual},
  author={Althoff, Matthias and Kochdumper, Niklas},
  journal={TU Munich},
  volume={85748},
  year={2016}
}

@inproceedings{michaux2023rdf,
  title={{Reachability-based Trajectory Design with Neural Implicit Safety Constraints}},
  author={Jonathan B Michaux AND Yong Seok Kwon AND Qingyi Chen AND Ram Vasudevan},
  booktitle={Proceedings of Robotics: Science and Systems},
  year={2023},
  doi={10.15607/RSS.2023.XIX.062}
}

@inproceedings{
    hadjiloizou2022Formal,
    title = {Formal Verification of Linear Temporal Logic Specifications Using Hybrid Zonotope-Based Reachability Analysis},
    author = {Hadjiloizou, Loizos and Jiang, Frank J. and Alanwar, Amr and Johansson, Karl H.},
    booktitle = {IEEE European Control Conference},
    year = {2024}
}

@article{raimondo2016closed,
  title={Closed-loop input design for guaranteed fault diagnosis using set-valued observers},
  author={Raimondo, Davide M and Marseglia, Giuseppe Roberto and Braatz, Richard D and Scott, Joseph K},
  journal={Automatica},
  volume={74},
  pages={107--117},
  year={2016},
  publisher={Elsevier}
}

@article{andrade2024tube,
  title={Tube-based model predictive control based on constrained zonotopes},
  author={Andrade, Richard and Normey-Rico, Julio E and Raffo, Guilherme V},
  journal={IEEE Access},
  year={2024},
}

@book{demmel1997applied,
  title={Applied Numerical Linear Algebra},
  author={Demmel, James W},
  year={1997},
  publisher={SIAM}
}

@inproceedings{vinod2024pycvxset,
  title={pycvxset: A Python package for convex set manipulation},
  author={Vinod, Abraham P},
  booktitle={IEEE American Control Conference},
  year={2025}
}

@article{schupp2022recent,
  title={Recent developments in theory and tool support for hybrid systems verification with {HyPro}},
  author={Schupp, Stefan and {\'A}brah{\'a}m, Erika and Ebert, Tristan},
  journal={Information and Computation},
  volume={289},
  pages={104945},
  year={2022},
  publisher={Elsevier}
}

@inproceedings{bogomolov2019juliareach,
  title={{JuliaReach}: a toolbox for set-based reachability},
  author={Bogomolov, Sergiy and Forets, Marcelo and Frehse, Goran and Potomkin, Kostiantyn and Schilling, Christian},
  booktitle={Proceedings of the 22nd ACM International Conference on Hybrid Systems: Computation and Control},
  pages={39--44},
  year={2019}
}

@article{huang2019dp,
  title={{DP-ADMM}: {ADMM}-based distributed learning with differential privacy},
  author={Huang, Zonghao and Hu, Rui and Guo, Yuanxiong and Chan-Tin, Eric and Gong, Yanmin},
  journal={IEEE Transactions on Information Forensics and Security},
  volume={15},
  pages={1002--1012},
  year={2019},
}

@article{elgabli2020q,
  title={{Q-GADMM}: Quantized group {ADMM} for communication efficient decentralized machine learning},
  author={Elgabli, Anis and Park, Jihong and Bedi, Amrit Singh and Issaid, Chaouki Ben and Bennis, Mehdi and Aggarwal, Vaneet},
  journal={IEEE Transactions on Communications},
  volume={69},
  number={1},
  pages={164--181},
  year={2020},
}

@inproceedings{zhang2022safety,
  title={Safety verification of neural feedback systems based on constrained zonotopes},
  author={Zhang, Yuhao and Xu, Xiangru},
  booktitle={IEEE 61st Conference on Decision and Control},
  pages={2737--2744},
  year={2022}
}

@article{rego2025novel,
  title={A Novel and Efficient Order Reduction for Both Constrained Convex Generators and Constrained Zonotopes},
  author={Rego, Francisco and Silvestre, Daniel},
  journal={IEEE Transactions on Automatic Control},
  year={2025},
}

@article{rego2020guaranteed,
  title={Guaranteed methods based on constrained zonotopes for set-valued state estimation of nonlinear discrete-time systems},
  author={Rego, Brenner S and Raffo, Guilherme V and Scott, Joseph K and Raimondo, Davide M},
  journal={Automatica},
  volume={111},
  pages={108614},
  year={2020},
  publisher={Elsevier}
}

@inproceedings{thompson2025mixed,
  title={Mixed-Integer Moving Horizon Estimation for Terrain-Aided Navigation Using Hybrid Zonotopes},
  author={Thompson, Andrew F and Robbins, Joshua A and Boler, Matthew E and Pangborn, Herschel C},
  booktitle={IEEE/ION Position, Location and Navigation Symposium (PLANS)},
  pages={815--820},
  year={2025}
}

@article{althoff2014online,
  title={Online verification of automated road vehicles using reachability analysis},
  author={Althoff, Matthias and Dolan, John M},
  journal={IEEE Transactions on Robotics},
  volume={30},
  number={4},
  pages={903--918},
  year={2014}
}

@inproceedings{rego2025zeta,
  title={{ZETA}: a library for {Zonotope-based EsTimation and fAult} diagnosis of discrete-time systems},
  author={Rego, Brenner S and Scott, Joseph K and Raimondo, Davide M and Terra, Marco H and Raffo, Guilherme V},
  booktitle={2025 IEEE 64th Conference on Decision and Control (CDC)},
  pages={2420--2427},
  year={2025}
}

@article{seo2022real,
  title={Real-time robust receding horizon planning using {Hamilton--Jacobi} reachability analysis},
  author={Seo, Hoseong and Lee, Donggun and Son, Clark Youngdong and Jang, Inkyu and Tomlin, Claire J and Kim, H Jin},
  journal={IEEE Transactions on Robotics},
  volume={39},
  number={1},
  pages={90--109},
  year={2022}
}

@article{bravo2006robust,
  title={Robust {MPC} of constrained discrete-time nonlinear systems based on approximated reachable sets},
  author={Bravo, Jos{\'e} Manuel and Alamo, Teodoro and Camacho, Eduardo F},
  journal={Automatica},
  volume={42},
  number={10},
  pages={1745--1751},
  year={2006},
  publisher={Elsevier}
}

@inproceedings{schurmann2018reachset,
  title={Reachset model predictive control for disturbed nonlinear systems},
  author={Sch{\"u}rmann, Bastian and Kochdumper, Niklas and Althoff, Matthias},
  booktitle={IEEE Conference on Decision and Control (CDC)},
  pages={3463--3470},
  year={2018}
}

@article{liu2024refine,
  title={{REFINE}: Reachability-based trajectory design using robust feedback linearization and zonotopes},
  author={Liu, Jinsun and Shao, Yifei Simon and Lymburner, Lucas and Qin, Hansen and Kaushik, Vishrut and Trang, Lena and Wang, Ruiyang and Ivanovic, Vladimir and Tseng, H Eric and Vasudevan, Ram},
  journal={IEEE Transactions on Robotics},
  volume={40},
  pages={2060--2080},
  year={2024}
}

@article{alamo2005guaranteed,
  title={Guaranteed state estimation by zonotopes},
  author={Alamo, Teodoro and Bravo, Jos{\'e} Manuel and Camacho, Eduardo F},
  journal={Automatica},
  volume={41},
  number={6},
  pages={1035--1043},
  year={2005},
  publisher={Elsevier}
}

@inproceedings{sadraddini2019linear,
  title={Linear encodings for polytope containment problems},
  author={Sadraddini, Sadra and Tedrake, Russ},
  booktitle={2019 IEEE 58th conference on decision and control},
  pages={4367--4372},
  year={2019}
}

@article{kousik2022ellipsotopes,
  title={Ellipsotopes: Uniting ellipsoids and zonotopes for reachability analysis and fault detection},
  author={Kousik, Shreyas and Dai, Adam and Gao, Grace Xingxin},
  journal={IEEE Transactions on Automatic Control},
  volume={68},
  number={6},
  pages={3440--3452},
  year={2022}
}

@article{silvestre2021constrained,
  title={Constrained convex generators: A tool suitable for set-based estimation with range and bearing measurements},
  author={Silvestre, Daniel},
  journal={IEEE Control Systems Letters},
  volume={6},
  pages={1610--1615},
  year={2021}
}

@phdthesis{bird2022hybrid,
  title={Hybrid zonotopes: A mixed-integer set representation for the analysis of hybrid systems},
  author={Bird, Trevor J},
  year={2022},
  school={Purdue University}
}
\bibliographystyle{ieeetr}

\end{document}